\documentclass[useAMS,usenatbib]{mn2e}
\usepackage[T1]{fontenc}
\usepackage{aecompl}
\usepackage{Times}
\usepackage{graphicx}
\usepackage{supertabular,multicol}
\usepackage{threeparttable}

\newcommand{\molh}{H$_2$}

\newcommand{\HPAH}{S(3)/PAH[11.3$\mu$m]}

\newcommand{\spi}{{\it Spitzer}}
\newcommand{\oiii}{[O {\sc iii}]$\lambda$5007\AA}
\newcommand{\feii}{[Fe {\sc ii}]}
\newcommand{\feiiMIR}{[Fe {\sc ii}]$\lambda$5.34\micron}
\newcommand{\nev}{[Ne {\sc v}]$\lambda$14.32\micron}
\newcommand{\neii}{[Ne {\sc ii}]$\lambda$12.81\micron}
\newcommand{\oi}{[O {\sc i}]$\lambda$6300\AA}
\newcommand{\oiv}{[O {\sc iv}]$\lambda$25.89\AA}
\newcommand{\feiiNIR}{[Fe {\sc ii}]$\lambda$1.644\micron}
\newcommand{\feiiPa}{[Fe {\sc ii}]$\lambda$1.644\micron/Pa$\alpha$}
\newcommand{\oir}{[O {\sc i}]$\lambda$6300\AA/H$\alpha$}
\newcommand{\nii}{[N {\sc ii}]$\lambda$6583\AA}
\newcommand{\niir}{[N {\sc ii}]$\lambda$6583\AA/H$\alpha$}
\newcommand{\sii}{[S {\sc ii}]$\lambda\lambda$6716,6731\AA}
\newcommand{\siir}{[S {\sc ii}]$\lambda\lambda$6716,6731\AA/H$\alpha$}
\newcommand{\texc}{$T_{\rm exc}$}

\title{Warm molecular hydrogen in outflows from Ultraluminous Infrared Galaxies}

\author[Hill \& Zakamska]{Matthew J. Hill$^1$\thanks{E-mail: mhill@pha.jhu.edu},
Nadia L. Zakamska$^1$\thanks{E-mail: zakamska@pha.jhu.edu}\\
\\
$^{1}$Department of Physics \& Astronomy, Johns Hopkins University, 3400 N. Charles St., Baltimore, MD 21218, USA \\
}

\begin{document}
\maketitle
\date{MNRAS, in review}

\pagerange{\pageref{firstpage}--\pageref{lastpage}} \pubyear{2013}

\maketitle

\label{firstpage}

\begin{abstract}
Ultraluminous infrared galaxies (ULIRGs) show on average three times more emission in the rotational transitions of molecular hydrogen than expected based on their star formation rates. Using \spi\ archival data we investigate the origin of excess warm \molh\ emission in 115 ULIRGs of the IRAS 1 Jy sample. We find a strong correlation between \molh\ and \feii\ line luminosities, suggesting that excess \molh\ is produced in shocks propagating within neutral or partially ionized medium. This view is supported by the correlations between \molh\ and optical line ratios diagnostic of such shocks. The galaxies powered by star formation and those powered by active nuclei follow the same relationship between \molh\ and \feii, with emission line width being the major difference between these classes ($\sim 500$ and $\sim 1000$ km/sec, respectively). We conclude that excess \molh\ emission is produced as the supernovae and active nuclei drive outflows into the neutral interstellar medium of the ULIRGs. A weak positive correlation between \molh\ and the length of the tidal tails indicates that these outflows are more likely to be encountered in more advanced mergers, but there is no evidence for excess \molh\ produced as a result of the collision shocks during the final coalescence.
\end{abstract}


\section{Introduction}
\label{Introduction}

Although the amount of molecular hydrogen \molh\ in gas-rich galaxies can reach $>10^{10}M_{\odot}$ \citep{solo97}, the rotational transitions of \molh\ in the mid-infrared and the ro-vibrational transitions in the near-infrared are relatively weak. One reason is that the rotational transitions are quantum-mechanically forbidden; another is that most molecular gas is at a temperature of just a few tens of K \citep{drai11}, much smaller than the temperature required to populate even the lowest excited levels of \molh\ (e.g., $E_{J=2}=k_B\times 518$ K). Thus the \molh\ we observe in the mid-infrared is ``warm'' molecular hydrogen which represents only a small fraction of the total \molh\ present in the galaxy \citep{higd06}.

Emission of warm \molh\ in star-forming galaxies is a minor bi-product of star formation, contributing only a small fraction ($<$ a few $\times 10^{-4}$) of the total infrared luminosity of the galaxy \citep{rous07}. The dominant source of excitation of the observed emission of molecular hydrogen \molh\ is likely in photo-dissociation regions around young stars \citep{shul82, holl97}. The strong correlation between the mid-infrared \molh\ emission and the star formation rate over several orders of magnitude in galaxy luminosity \citep{rous07} supports this view. In this case, the dominant process of \molh\ excitation is due to ultra-violet photons of newly formed massive stars, which either pump \molh\ molecules straight into excited electronic states or heat the surrounding dust which then heats the gas.

In an unexpected challenge to this picture, multiple \spi\ observations uncovered extragalactic objects with unusually high rotational \molh\ luminosities in which little or no star formation was seen \citep{appl06, egam06, ogle07, john07}. Within this sample, there may be several distinct physical processes responsible for \molh\ excitation. In the first class are objects associated with large-scale shocks, either due to relativistic jets \citep{ogle10} or galaxy collisions \citep{appl06, ogle07, guil09} or infall of intra-cluster gas onto the central cluster galaxy \citep{egam06}. In the second class are intra-cluster filaments where \molh\ is seemingly excited by the deposition of cosmic rays into molecular gas \citep{john07, ferl08}. Furthermore, the molecular gas in central cluster galaxies may be heated by electrons from the hot cluster atmosphere \citep{dona11}.

The phenomenon of excess emission from warm molecular hydrogen ($T\sim$ a few hundred K) has now been observed in a wide range of objects and environments, from spiral galaxies \citep{beir09} to active nuclei \citep{dasy11} to Galactic translucent clouds \citep{inga11}. Not only does the molecular gas at these relatively warm temperatures constitute an unusual phase of the interstellar medium, but this gas is typically found out of the dynamic equilibrium with the rest of the galaxy and is often a signature of feedback, whether due to star formation \citep{rous07, veil09b} or accretion onto a supermassive black hole \citep{nesv10, ogle10, dasy11}.

The amount of \molh\ emission seen from ultraluminous infrared galaxies (ULIRGs) is also in excess of that expected from star formation alone \citep{zaka10}, by about a factor of three or more. In this paper we investigate the nature of the \molh\ emission in these objects. Our sample and data are described in Sec. \ref{sec:sample}. In Sec. \ref{sec:morph}, we discuss the relationship between the warm molecular gas emission and the global morphological properties of the galaxies. In Sec. \ref{sec:phases} we discuss the relationships between different phases of the gas: molecular, neutral and warm ionized. We present our conclusions in Sec. \ref{sec:conclusions}. We use an $h=0.7$, $\Omega_m=0.3$, $\Omega_{\Lambda}=0.7$ cosmology throughout this paper. Wavelengths of emission lines are given in air, except in Section \ref{sec_opt_spec} where we discuss fits to Sloan Digital Sky Survey spectra which use vacuum wavelengths. In these cases, the vacuum wavelengths are marked with `v'.

To evaluate the significance of weak correlations, we use the Spearman rank test, and to evaluate the similarity of distributions we use the Kolmogorov-Smirnov (KS) test. For both tests we report the probability of the null hypothesis as $p_{\rm S}$ or $p_{\rm KS}$ -- in the first case, the probability that the two sets of data are uncorrelated and in the second case, that the two sets are drawn from the same distribution. Small values of $p_{\rm S}$ indicate the presence of a statistically significant correlation and small values of $p_{\rm KS}$ indicate a statistically significant difference between two distributions.

\section{Sample, data and measurements}
\label{sec:sample}

\subsection{Sample selection}

We start with the Infrared Astronomical Satellite (IRAS) 1 Jy sample of 118 ULIRGs \citep{kim98a}. These sources are selected from the IRAS Faint Source Catalog to have flux $F_{\nu}(60 \mu m) > 1$ Jy and are constrained to be at high Galactic latitudes $|b| > 30^{\circ}$ and declinations $\delta > -40^{\circ}$. The objects cover a redshift range $z = 0.02-0.27$ and the infrared luminosity range $\log {(L_{\rm IR}/L_{\sun})} = 12.00-12.90$. Because the spatial resolution of IRAS is $\ga 1$\arcmin, a single IRAS source can be composed of two or more individual galaxies which are not resolved individually in the IRAS catalog.

We search \spi\ Legacy Archive for low-resolution Infrared Spectrograph (IRS, \citealt{houc04}) data within 60\arcsec\ of the nominal position of each target. We use the enhanced spectroscopic data products supplied by the \spi\ Science Center (SSC), in which the spectrum has been extracted, background-subtracted and flux-calibrated using the standard spectroscopic pipeline (v. S18.18.0). In many cases, multiple separate visits to the same target are available in the archive. We combine all such spectra into one using error-weighted averages. The close interacting pairs present a more difficult issue. In some such cases only one component of the pair has a spectrum in the archive, and sometimes both do. If the components are closer than 4\arcsec\ on the sky -- comparable to or smaller than the IRS slit size -- we consider them to be unresolved by the IRS and combine such spectra into one. If they are further apart, we consider them as separate sources. We remove the 5 objects where only the short-wavelength orders are available.

Our final sample is defined by the 115 distinct \spi\ spectra (some of which are individual components within a single IRAS source) with both short-low (SL) observations with wavelength coverage $5.2-14.5\micron$ and slit dimensions 4.7\arcsec$\times$11.3\arcsec\ and long-low (LL) observations with wavelength coverage $14.0-38.0\micron$ and slit dimensions 11.1\arcsec$\times$22.3\arcsec. We present the sample in Table \ref{tab:sample} using the IRS pointing information for precise identification of the sources (keywords RA\_RQST and DEC\_RQST in the SSC spectroscopic reductions). There is often a discrepancy between the flux normalization in the LL orders and the SL orders, in that LL fluxes are higher. This is most likely due to the difference in apertures: LL aperture is larger and lets in more flux for extended objects. In such cases we apply a multiplicative factor ($\le1$) to the LL spectrum to ensure that both the flux and the spectral slope are continuous through 14\micron\ where LL and SL spectra overlap. The factors are listed in Table \ref{tab:sample}.  The final combined and renormalized spectra are available as ASCII files in the online edition of the journal.

We compare our final spectra with those from \citet{will11} who used a custom reduction procedure different from the standard SSC pipeline. There are 30 overlapping galaxies between their sample and ours, which we use to check the accuracy of the spectrophotometry of our sources. For every source, we normalize the SSC spectrum and the \citet{will11} spectrum to the same value at 22\micron, divide the two one by the other and calculate the median values of this ratio in each of the four orders involved (SL1, SL2, LL1, LL2) and find that the relative spectrophotometry between the two sets of spectra agrees within 20\% (standard deviation). Thus a line ratio calculated from two different orders in the \spi\ spectrum should be regarded as having a systematic uncertainty of $\la 0.1$ dex (and much smaller for a line ratio computed within a single order). 

\begin{table}
 \caption{ULIRGs from \emph{IRAS} 1 Jy Sample that have \spi\ spectra}
 \label{tab:sample}
 \begin{tabular}{llllc}
  \hline
  \multicolumn{1}{c}{IRAS Name} &
  \multicolumn{1}{c}{Redshift} &
  \multicolumn{1}{c}{R.A.} &
  \multicolumn{1}{c}{Decl.} &
  \multicolumn{1}{c}{Mult. Factor} \\
  \multicolumn{1}{c}{(1)} &
  \multicolumn{1}{c}{(2)} &
  \multicolumn{1}{c}{(3)} &
  \multicolumn{1}{c}{(4)} &
  \multicolumn{1}{c}{(5)} \\
  \hline
IRAS 00091-0738  &   0.118 &   2.930210 &  -7.368560 & 0.878 \\
IRAS 00188-0856  &   0.128 &   5.360330 &  -8.657530 & 1.000 \\
... & ... & ... & ... & ... \\
IRAS 01003-2238    & 0.118 &  15.708080 & -22.365917 & 0.956 \\
IRAS 01166-0844:NW & 0.118 &  19.781460 &  -8.485940 & 0.220 \\
IRAS 01166-0844:SE & 0.118 &  19.782673 &  -8.486669 & 1.000 \\
... & ... & ... & ... & ... \\
  \hline
 \end{tabular}\\
 {\bf Notes.} --
 {\bf (1)} IRAS Name, supplemented by component designation if necessary using images from \citet{kim02} (e.g., IRAS 01166-0844:NW for the North-Western component of IRAS 01166-0844). 
 {\bf (2)} Redshift from \citet{kim98a}.
 {\bf (3)} Right Ascension of {\it Spitzer} pointing at J2000.00 epoch in degrees.
 {\bf (4)} Declination of {\it Spitzer} pointing at J2000.00 epoch in degrees.
 {\bf (5)} Multiplicative factor used to correct for aperture size difference in SL and LL slits. \\
  (A full version of this table is available in the online journal.)
\end{table}

\subsection{Analysis of \spi\ spectra}

\begin{figure*}
\centering
\includegraphics[scale=0.65]{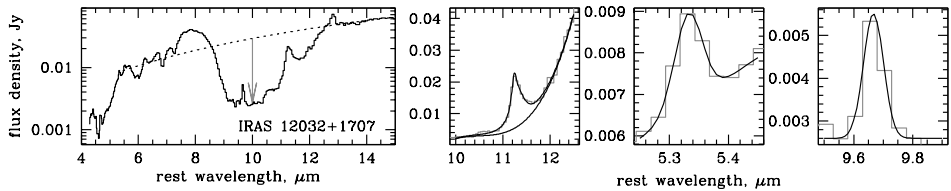}
\caption{{\bf Left:} Example IRS spectrum of one of the objects in our sample. The dashed line shows the interpolated power-law continuum and the arrow shows the measurement of the strength of the silicate feature, $\tau[9.7\micron]$. {\bf Right:} fits to the PAH[11.3\micron] feature, to the \feiiMIR\ line and to the S(3) line of molecular hydrogen.}\label{pic:fits}
\end{figure*}

The spectrum of rotational transitions of \molh\ is composed of transitions from rotational energy levels with quantum numbers $J+2$ to those with quantum numbers $J$; the corresponding emission features are denoted S($J$). We measure the fluxes of S(1) and S(3) molecular hydrogen emission lines by fitting a Gaussian profile with a quadratic continuum to a cropped section of the spectrum; the other rotational \molh\ lines are much weaker and are not detected in most spectra. To check the validity of our flux measurements we compare with measurements of a sample of ULIRGs by \citet{higd06}. There are 26 objects shared between the samples. In 20 of them, the measurements of S(1) and S(3) fluxes by \citet{higd06} and us agree within 1$\sigma$, whereas in the remaining 6 they agree within 3$\sigma$. We assume rest-frame centroids of the lines to be at 17.0346\micron\ and 9.6649\micron, and we use optical redshifts from \citet{kim98a} to find the observed wavelengths of the features. The extraction regions are set to 16.6-17.3\micron, 9.1-10.2\micron\ and 5.23-5.46\micron\ to avoid other strong emission features. We allow a slop in line centroid of 0.05\micron\ to account for a possible uncertainty in the redshift and in the IRS wavelength calibration. While the widths of the lines are not fixed in the fit, we find that they agree well with those expected due to the instrumental resolution \citep{smit07} which varies between $R=\lambda/\Delta\lambda=64-128$ as a function of the observed wavelength. Example fits are shown in Figure \ref{pic:fits}.

\begin{table*}
 \caption{Mid-Infrared Properties: \molh\ and \feii\ emission}
 \label{tab:midir1}
 \begin{tabular}{lllllllllll}
  \hline
  \multicolumn{1}{c}{IRAS Name} &
  \multicolumn{1}{c}{$L$[S(1)]} &
  \multicolumn{1}{c}{$\sigma$[S(1)]} &
  \multicolumn{1}{c}{$L$[S(3)]} &
  \multicolumn{1}{c}{$\sigma$[S(3)]} &
  \multicolumn{1}{c}{$T_{\rm exc}$} &
  \multicolumn{1}{c}{$\sigma$[$T_{\rm exc}$]} &
  \multicolumn{1}{c}{$L$[Fe {\sc ii}]} &
  \multicolumn{1}{c}{$\sigma$[Fe {\sc ii}]} &
  \multicolumn{1}{c}{EW PAH[6.2$\mu$m]} &
  \multicolumn{1}{c}{$\sigma$(EW) PAH[6.2$\mu$m]} \\
  \multicolumn{1}{c}{(1)} &
  \multicolumn{1}{c}{(2)} &
  \multicolumn{1}{c}{(3)} &
  \multicolumn{1}{c}{(4)} &
  \multicolumn{1}{c}{(5)} &
  \multicolumn{1}{c}{(6)} &
  \multicolumn{1}{c}{(7)} &
  \multicolumn{1}{c}{(8)} &
  \multicolumn{1}{c}{(9)} &
  \multicolumn{1}{c}{(10)} &
  \multicolumn{1}{c}{(11)} \\
  \hline
IRAS 00091-0738 & 37.70 & 3.682 & 10.14 & 3.140 & 324 & 23 & 11.46 & 4.277 & 0.202 & 0.02460\\
IRAS 00188-0856 & 31.41 & 6.645 & 20.61 & 3.601 & 402 & 30 & 8.287 & 5.429 &  0.246 & 0.00986\\
IRAS 00397-1312 & \ldots & \ldots & 98.40 & 15.37 & 411 & 18 & 106.9 & 74.66 & 0.053 & 0.00134\\
IRAS 00456-2904 & 28.36 & 3.001 & 19.66 & 1.709 & 408 & 15 & 16.84 & 3.287 & 1.399 & 0.8211\\
IRAS 00482-2720 & 67.27 & 9.092 & 11.96 & 3.650 & 297 & 20 & \ldots & \ldots & 1.057 & 0.2809\\
  \hline
 \end{tabular}\\
 {\bf Notes.} --
 {\bf (1)} Object IRAS Name.
 {\bf (2,3)} Luminosity of S(1) emission line in $10^{40}$ erg s$^{-1}$ and its uncertainty.
 {\bf (4,5)} Luminosity of S(3) emission line in $10^{40}$ erg s$^{-1}$ and its uncertainty.
 {\bf (6,7)} Excitation temperature derived from S(3)-S(1) transition in degrees K and its uncertainty.
 {\bf (8,9)} Luminosity of \feiiMIR\ in $10^{40}$ erg s$^{-1}$ and its uncertainty.
 {\bf (10,11)} Equivalent width of the PAH[6.2\micron] in microns and its uncertainty.\\
  (A full version of this table is available in the online journal.)
\end{table*}

The population of each energy level can be calculated from the measured flux of the transition:
\begin{equation}\label{pop}
  N_{J+2} = \frac{4\pi D_{L}^{2}F_{J}}{A_{J+2\rightarrow J}(E_{J+2} - E_{J})},
\end{equation}
where $F_{J}$ is the line flux, $D_{L}$ is the luminosity distance, and $A$ are the Einstein coefficients \citep{turn77}. Here $E_{J+2} - E_{J}$ is the energy of the transition (equal to the energy of each emitted photon), with energy levels given by \citet{hube79} and \citet{land91} as:
\begin{equation}\label{eng}
  E_{J} = 85.35K\cdot k_{B} J(J+1) - 0.068K\cdot k_{B} J^{2}(J+1)^{2},
\end{equation}
where $k_{B}$ is the Boltzmann constant. The excitation temperature is related to level populations via
\begin{equation}\label{temp}
  \frac{N_{J}}{g_{J}} \propto \exp{\left(-\frac{E_{J}}{k_{B}T_{\rm exc}}\right)},
\label{eq_texc}
\end{equation}
where $g_{J}$ is the degeneracy of the levels, with $g_{J} = 3(2J+1)$ for odd $J$ and $g_{J} =2J+1$ for even $J$. Thus, the excitation temperature can be calculated given any two or more line fluxes.

Normally a wide range of excitation temperatures is present, resulting in a non-linear relationship between $\ln(N_J/g_J)$ and $E_J$ \citep{armu06}. Typical two-temperature fits to this relationship in ULIRGs \citep{higd06} reveal a large mass ($\sim 10^9M_{\odot}$) of colder ($T\la 300$K) gas producing emission in low-$J$ transitions and a much lower mass ($\sim 10^6 M_{\odot}$) of warmer ($T\ga 1000$K) gas that dominates high-$J$ fluxes. But it is likely that in every object there is a distribution of gas masses as a function of temperature \citep{zaka10}. In what follows we utilize only the excitation temperature \texc\ between levels $J=3$ and $J=5$ measured using the S(1) and the S(3) transitions, with uncertainty in \texc\ derived through propagation of uncertainties of the S(1) and S(3) line fluxes following equation (\ref{eq_texc}). In a subsample of ULIRGs presented here, \citet{zaka10} derives excitation temperatures using S(1), S(2) and S(3) lines and finds a median temperature of 330 K, lwer than the median temperature of 406 K calculated using just the S(1) and S(3) lines. This is not surprising because including the S(2) line flux has the effect of giving more statistical weight to the lower-$J$ transitions whose flux is dominated by emission from lower temperature gas. For some of the objects in our sample it would be possible to construct more detailed excitation diagrams and multi-temperature fits, but in most cases because of the low signal-to-noise of the data we can measure only these two strongest \molh\ lines, and thus we restrict ourselves to this simple measure. Physically, the \texc\ that we measure may be interpreted in terms of the slope of the mass-temperature distribution, or in terms of the the ratio of mass of the warmer component to that of the cooler component. 

We use the emission of polycyclic aromatic hydrocarbons (PAH; \citealt{alla89}) as a measure of star formation \citep{rous01, dale02}. PAH fluxes are calculated by cutting out a $\la3\micron$-wide part of the spectrum, excluding other features in that wavelength range, and then modeling this cut-out using a polynomial continuum and Drude (damped harmonic oscillator) profiles with profile shapes and widths taken from \citet{smit07}. For PAH complexes, such as those at 11.3\micron\ and at 7.7\micron, the relative amplitudes of the components within the complex are fixed to their ratios in the template spectrum of normal star-forming galaxies \citep{smit07}. For example, within the 11.3\micron\ complex the amplitude ratio of the 11.23\micron\ and 11.33\micron\ components is fixed to 1.25:1. Depending on the model for the local continuum, from a constant to  a cubic polynomial, the number of fit parameters varies from two to five, with the amplitude being the only parameter that describes the intensity of the PAH feature (since the functional shape of the feature remains fixed). Example fit is shown in Figure \ref{pic:fits}. The quality of fits to PAH features is typically very good, and most are detected in almost all objects with high confidence.

\begin{table*}
 \caption{Mid-Infrared Properties: Si absorption and PAH emission}
 \label{tab:midir2}
 \begin{tabular}{lllllllllll}
  \hline
  \multicolumn{1}{c}{IRAS Name} &
  \multicolumn{1}{c}{$\tau$[9.7$\mu$m]} &
  \multicolumn{1}{c}{$L$[6.2$\mu$m]} &
  \multicolumn{1}{c}{$\sigma$[6.2$\mu$m]} &
  \multicolumn{1}{c}{$L$[7.7$\mu$m]} &
  \multicolumn{1}{c}{$\sigma$[7.7$\mu$m]} &
  \multicolumn{1}{c}{$L$[8.5$\mu$m]} &
  \multicolumn{1}{c}{$\sigma$[8.5$\mu$m]} &
  \multicolumn{1}{c}{$L$[11.3$\mu$m]} &
  \multicolumn{1}{c}{$\sigma$[11.3$\mu$m]} \\
  \multicolumn{1}{c}{(1)} &
  \multicolumn{1}{c}{(2)} &
  \multicolumn{1}{c}{(3)} &
  \multicolumn{1}{c}{(4)} &
  \multicolumn{1}{c}{(5)} &
  \multicolumn{1}{c}{(6)} &
  \multicolumn{1}{c}{(7)} &
  \multicolumn{1}{c}{(8)} &
  \multicolumn{1}{c}{(9)} &
  \multicolumn{1}{c}{(10)}\\
  \hline
IRAS 00091-0738 &    2.70 & 4.92 & 0.69 & 39.28 & 5.59 & 0.00 & 1.13 & 4.93 & 0.54 \\
IRAS 00188-0856 &    2.53 & 10.34 & 0.82 & 60.74 & 10.25 & 16.75 & 1.15 & 8.62 & 0.56 \\
IRAS 00397-1312 &    2.51 & 76.35 & 6.86 & 252.40 & 30.09 & 0.00 & 10.03 & 29.41 & 1.56 \\
IRAS 00456-2904 &    0.82 & 18.42 & 1.13 & 68.32 & 2.19 & 22.51 & 1.17 & 12.52 & 0.58 \\
IRAS 00482-2720 &    2.13 & 5.00 & 0.55 & 16.30 & 1.83 & 4.53 & 1.10 & 2.58 & 0.54 \\
  \hline
 \end{tabular}\\
 {\bf Notes.} --
 {\bf (1)} Object IRAS Name.
 {\bf (2)} Strength of the silicate feature at 9.7 $\mu$m.
 {\bf (3,4)} Luminosity of PAH[6.2$\mu$m] emission feature in $10^{42}$ erg s$^{-1}$ and its uncertainty.
 {\bf (5,6)} Luminosity of PAH[7.7$\mu$m] emission feature in $10^{42}$ erg s$^{-1}$ and its uncertainty.
 {\bf (7,8)} Luminosity of PAH[8.5$\mu$m] emission feature in $10^{42}$ erg s$^{-1}$ and its uncertainty.
 {\bf (9,10)} Luminosity of PAH[11.3$\mu$m] emission feature in $10^{42}$ erg s$^{-1}$ and its uncertainty.
  (A full version of this table is available in the online journal.)

\end{table*}

The powering source of ULIRGs (star formation and/or accretion onto a supermassive black hole) is often hidden behind large amounts of dust and gas, which may be optically thick even at mid-infrared wavelengths. We parametrize the dust obscuration using the apparent strength of silicate absorption defined as $\tau[9.7\micron]=-\ln(f_{\rm obs}[9.7\micron]/f_{\rm cont}[9.7\micron])$, where $f_{\rm obs}$ is the observed flux density and $f_{\rm cont}$ is the guess for the local continuum. To estimate $f_{\rm cont}$, we average emission in the wavelength ranges $5.3-5.6\micron$ and $13.85-14.15\micron$ and then interpolate between these points using a power-law. In the calculation of $f_{\rm obs}$, the S(3) line is excluded and the continuum within the absorption trough is fitted with a second order polynomial. Negative values of $\tau[9.7\micron]$ indicate silicate emission, while positive values indicate absorption, with $\tau[9.7\micron]\simeq 3.5$ for the most absorbed sources. This method is almost identical to that used by \citet{spoo07}; the minor difference at high optical depths is that we do not use a continuum point at 7.7\micron\ even in the cases of weak PAH emission. The apparent strength of Si absorption defined this way is closely related to, but not identical to the optical depth of Si dust absorption; depending on the continuum opacity of the dust at these wavelengths, the actual optical depth is $\simeq (1.2-1.5) \times$ the apparent strength of the Si feature.

Many other emission features are present in \spi\ spectra. The one of particular interest to us is the \feiiMIR\ emission line which we measure using the same method (single Gaussian fit) as the \molh\ emission. Measurements made from \spi\ spectra are given in Tables \ref{tab:midir1} and \ref{tab:midir2}.

\subsection{Optical and near-infrared morphologies}

The vast majority of ULIRGs show signs of on-going or recent merger activity \citep{sand96} and are likely fueled by star formation in major gas-rich mergers \citep{dasy06a}. We use optical and near-infrared data to determine the stage of the merger and to provide quantitative measures along the merger sequence. Because of projection effects, such measures are only approximate.

A number of useful measurements of the IRAS 1 Jy sample are available in the literature. High-quality seeing-limited optical (\emph{R-}band) and near-infrared (\emph{K'}-band) images of IRAS 1 Jy sample are presented by \citet{kim02}, and measurements of nuclear separation and tidal tail length and merger stage classification are presented by \citet{veil02}. Nuclear separations can be measured down to the seeing limit of approximately $1\arcsec \sim3$ kpc. The lengths of tidal tails are recorded by measuring along each tail down to a constant surface brightness of 24 mag/arcsec$^2$ in $R$-band images.

\begin{table}
 \caption{Morphological and spectral classification}
 \label{tab:morph1}
 \begin{tabular}{lrrll}
  \hline
  \multicolumn{1}{c}{IRAS Name} &
  \multicolumn{1}{c}{NS$^a$} &
  \multicolumn{1}{c}{TL$^a$} &
  \multicolumn{1}{c}{MS$^a$} &
  \multicolumn{1}{c}{ST$^b$} \\
  \multicolumn{1}{c}{(1)} &
  \multicolumn{1}{c}{(2)} &
  \multicolumn{1}{c}{(3)} &
  \multicolumn{1}{c}{(4)} &
  \multicolumn{1}{c}{(5)} \\
  \hline
IRAS 00091-0738 &    2.1 & 12.0 & pre-merger & cp \\
IRAS 00188-0856 &    0.0 & 0.0  &  old merger & S2 \\
IRAS 00397-1312 &    0.0 & 0.0  &  old merger & cp \\
IRAS 00456-2904 &   20.7 & 9.0 & pre-merger &  cp \\
IRAS 00482-2720 &    6.7 & 15.0 & pre-merger &  cp \\
  \hline
 \end{tabular}\\
 {\bf Notes.} --
 {\bf (1)} Object IRAS Name.
 {\bf (2)} Projected nuclear separation measured in kpc.
 {\bf (3)} Length of tidal tails measured in kpc.
 {\bf (4)} Merger stage.
 {\bf (5)} Spectral type.\\
 $^a$ \citet{veil02}\\
 $^b$ \citet{yuan10}\\
  (A full version of this table is available in the online journal.)

\end{table}

\citet{veil02} use these measurements as well as their imaging observations \citep{kim02} to classify the merger stage of each object in the IRAS 1 Jy sample. These authors find that the combination of morphological features and the high infrared luminosities of ULIRGs indicate that the vast majority of the merging galaxies in these systems have already made the first close approach. This is likely a selection effect: only systems with the highest star formation rate have high enough luminosities to be classified as ULIRGs, and only the systems past their first approach have had enough of a potential perturbation to funnel enough gas into the centres of participating galaxies \citep{barn91,barn96,miho94,miho96}. Thus, \citet{veil02} place almost all ULIRGs into the III, IV, and V merger classes of their classification scheme, skipping I and II (which correspond to before and during the closest approach). Following \citet{veil02}, we refer to classes III, IV and V as Pre-merger, Merger and Old Mergers respectively.

Pre-merger systems are characterized by having two distinguishable nuclei as well as strong tidal features (which tend to develop after the first closest approach, \citealt{toom72}). Merger systems have a single detectable nucleus as well as strong tidal features, as well as an extended nuclear region. Old Mergers differ from Merger in that show no clear signs of tidal feature, but still have a disturbed central morphology indicative of a past merger event. In our sample of 115 systems, 49 have two distinguishable nuclei (Pre-Mergers), 48 are in the Merger stage and 12 are Old Mergers (Table \ref{tab:morph1}). There are also 5 galaxies in triple systems and 1 isolated non-interacting galaxy.

\subsection{Optical and near-infrared spectroscopic data}
\label{sec_opt_spec}

Optical and near-infrared spectra of ULIRGs show prominent emission lines produced in ionized gas. We use optical classifications by \citet{yuan10} to determine the dominant source of gas ionization. The first step is identification of objects with broad emission lines (FWHM$\sim$ a few thousand km sec$^{-1}$ for H$\beta$ and other permitted transitions), which are classified as optically unobscured active galactic nuclei (AGNs). For objects that have only narrow emission lines, standard diagnostic diagrams are used to determine the source of the ionization. Among our sample, 9 objects are classified as type 1 (broad-line) AGNs, 28 objects are type 2s (narrow-line, but with line ratios characteristic of photo-ionization by an obscured AGN), 13 objects are purely star-forming, 53 are star formation / AGN composites and 3 are low-ionization narrow emission-line region galaxies; 9 objects are left unclassified (Table \ref{tab:morph1}).

Furthermore, we use optical line flux measurements by \citet{veil99} which are obtained by fitting a Gaussian profile with a first-order polynomial continuum to host-subtracted spectra. We use these to construct diagnostic line ratios \oir, \niir\ and \siir\ and to calculate the Balmer decrement H$\alpha$/H$\beta$. These measurements are available for $>$90 per cent of the objects in our sample (Table \ref{tab:morph2}).

\begin{table*}
 \caption{Optical spectroscopic measurements from \citet{veil99}}
 \label{tab:morph2}
 \begin{tabular}{lrrllll}
  \hline
  \multicolumn{1}{c}{IRAS Name} &
  \multicolumn{1}{c}{$\log\frac{{\rm H}\alpha}{{\rm H}\beta}$} &
  \multicolumn{1}{c}{FWHM [O {\sc iii}]} &
  \multicolumn{1}{c}{$\log\frac{\rm[O~{\scriptscriptstyle III}]}{{\rm H}\beta}$} &
  \multicolumn{1}{c}{$\log\frac{\rm[N~{\scriptscriptstyle II}]}{{\rm H}\alpha}$} &
  \multicolumn{1}{c}{$\log\frac{\rm[S~{\scriptscriptstyle II}]}{{\rm H}\alpha}$} &
  \multicolumn{1}{c}{$\log\frac{\rm[O~{\scriptscriptstyle I}]}{{\rm H}\alpha}$} \\
  \multicolumn{1}{c}{(1)} &
  \multicolumn{1}{c}{(2)} &
  \multicolumn{1}{c}{(3)} &
  \multicolumn{1}{c}{(4)} &
  \multicolumn{1}{c}{(5)} &
  \multicolumn{1}{c}{(6)} &
  \multicolumn{1}{c}{(7)} \\
  \hline
IRAS 00091-0738 & 0.91 & \ldots  & -0.37 & -0.32 & -0.37 & -1.54 \\
IRAS 00188-0856 & 1.17 & \ldots & -0.44 & 0.13 & -0.44 & -0.93 \\
IRAS 00397-1312 & 0.86 & 440.0 & -0.53 & -0.35 & -0.53 & -1.28 \\
IRAS 00456-2904 & 0.80 & 340.0  & -0.47 & -0.30 & -0.47 & -1.41 \\
IRAS 00482-2720 & 0.97 & \ldots & -0.71 & -0.18 & -0.71 & -0.74 \\
  \hline
 \end{tabular}\\
 {\bf Notes.} --
 {\bf (1)} Object IRAS Name.
 {\bf (2)} Flux of H$\alpha$ relative to flux of H$\beta$.
 {\bf (3)} Full width at half maximum of \oiii\ in km/s.
 {\bf (4)} Logarithm of flux of \oiii\ relative to H$\beta$.
 {\bf (5)} Logarithm of flux of \nii\ relative to H$\alpha$.
 {\bf (6)} Logarithm of flux of the sum of \sii\ relative to H$\alpha$.
 {\bf (7)} Logarithm of flux of \oi\ relative to H$\alpha$.\\
  (A full version of this table is available in the online journal.)

\end{table*}

Additionally, we use \feiiNIR\ and Pa$\alpha$ fluxes from the near-infrared spectroscopic survey of ULIRGs by \citet{veil97}. Because of the weakness of the \feiiNIR\ emission line, the \feiiPa\ measurements are available only for 10 objects.

Finally, we search the spectroscopic database\footnote{http://mirror.sdss3.org/bulkSpectra} of the Sloan Digital Sky Survey \citep{york00} within 5\arcsec\ of the IRS positions of all our objects. Because some objects are found in close pairs, we examine all possible matches visually to ensure a proper correspondence between the galaxies with IRS spectra and the galaxies with SDSS spectra and end up with 36 objects that are listed in Table \ref{tab:opt}. Although most galaxies in our \spi\ sample are bright enough in the optical to be selected for the main galaxy survey of the SDSS \citep{stra02}, the SDSS observations cover a quarter of the sky whereas the \spi\ sample is selected over the entire sky; hence only a third of the objects are matched. Since the SDSS spectroscopic database operates on the vacuum wavelength scale, we use vacuum wavelengths for fits to SDSS data and mark with `v' all such cases in this section.

\begin{table*}
 \caption{Optical measurements from the SDSS spectra for a subsample of 36 ULIRGs}
 \label{tab:opt}
 \begin{tabular}{lllllll}
  \hline
  \multicolumn{1}{c}{IRAS Name} &
  \multicolumn{1}{c}{index} &
  \multicolumn{1}{c}{SDSS ID} &
  \multicolumn{1}{c}{$v_{\rm Na}$} &
  \multicolumn{1}{c}{EEW(NaD)} &
  \multicolumn{1}{c}{$v_{02}$([OIII])} &
  \multicolumn{1}{c}{$w_{90}$([OIII])} \\
  \multicolumn{1}{c}{(1)} &
  \multicolumn{1}{c}{(2)} &
  \multicolumn{1}{c}{(3)} &
  \multicolumn{1}{c}{(4)} &
  \multicolumn{1}{c}{(5)} &
  \multicolumn{1}{c}{(6)} &
  \multicolumn{1}{c}{(7)} \\
  \hline
IRAS 01166-0844:NW &  6 & 0660-52177-0563 &   70 &  2.50 &  -114 &   235 \\
IRAS 01572+0009    & 13 & 0403-51871-0550 &    0 &  0.00 & -1661 &  1845 \\
IRAS 03209-0806    & 17 & 0460-51924-0093 & -342 &  5.14 & -1284 &  1289 \\
IRAS 08201+2801    & 26 & 1267-52932-0383 & -120 &  1.68 &  -289 &   447 \\
IRAS 08559+1053    & 28 & 2575-54085-0125 &  179 &  5.46 &  -486 &   804 \\
 \hline
 \end{tabular}\\
 {\bf Notes.} --
 {\bf (1)} Object IRAS Name.
 {\bf (2)} 0-based index of the object in the \spi\ tables.
 {\bf (3)} SDSS spectroscopic identification (plate -- MJD of observations -- fiber).
 {\bf (4)} Velocity centroid of the Na D absorption feature relative to the stellar continuum, in km sec$^{-1}$.
 {\bf (5)} Estimated equivalent width of the Na D absorption feature, in \AA.
 {\bf (6)} Maximal blueshift of the [OIII]$\lambda$5007\AA\ emission line relative to the stellar continuum, in km sec$^{-1}$.
 {\bf (7)} Velocity width containing 90 per cent of the [OIII] line power, in km sec$^{-1}$. \\
  (A full version of this table is available in the online journal.)
\end{table*}

\begin{figure*}
\centering
\includegraphics[scale=0.70]{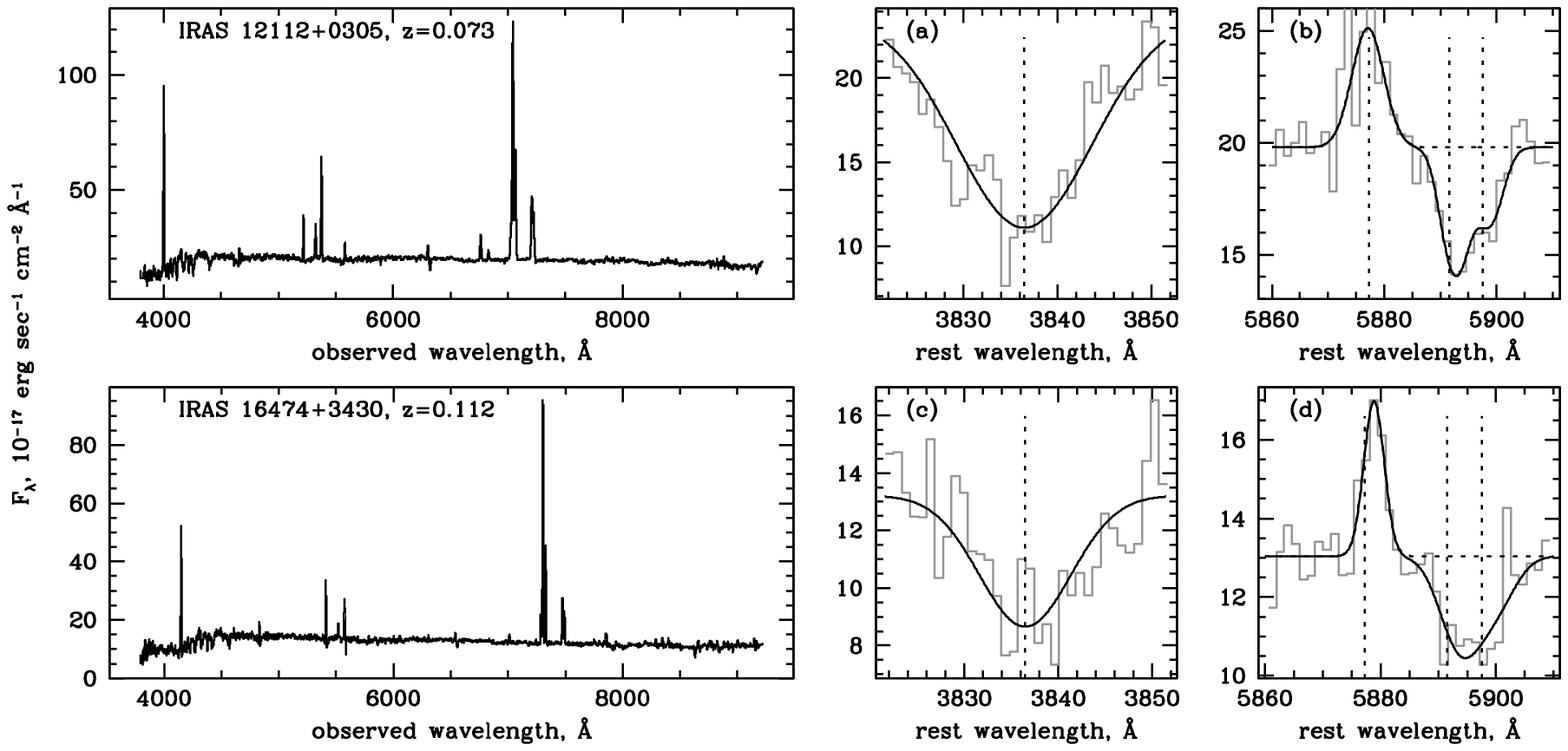}
\caption{{\bf Left:} Example optical spectra of two ULIRGs in our sample from the SDSS database. {\bf Middle and right:} Fits to stellar Balmer absorption features (panels a and c) and to the He I emission and Na D doublet absorption (panels b and d) in the same two objects (IRAS 12112+0306 and IRAS 16474+3430). Vertical dashed lines show the laboratory wavelengths of the features in question. The rest wavelengths are determined by finding accurate centroids of Balmer absorption features (in this case, H9 or H$\eta$ with vacuum wavelength 3936.47\AA); thus, in panels a and c the Gaussian fits are centered on their laboratory values by design. In the He I / Na D fits velocity offsets are allowed.}
\label{fig:opticalfits}
\end{figure*}

Because we are interested in the kinematics of the gas relative to the galaxy potential, we start by computing accurate redshifts of the stellar component of the galaxy. To this end, we use absorption features that are due to stellar photospheres. In particular, most spectra show high-order Balmer absorption lines (v3836.47\AA, v3798.98\AA, v3771.70\AA), presumably due to A stars common to these starbursting or post-starbursting objects. If these are not available, or if they are swamped with emission features of ionized interstellar gas, then we use Ca II v3934.78\AA, but we do not prefer it over the Balmer lines because it may be contaminated by interstellar absorption. When one or more of these absorption lines are detected, we fit them with Gaussians to determine the absorption line centroids, and use median averaging if more than one absorption line is available. In a few cases none of these or other common stellar absorption features are detected and we adopt the SDSS pipeline redshift. Example fits are shown in Figure \ref{fig:opticalfits}. Whenever we make kinematic measurements from the SDSS data, we use the accurate stellar redshifts as a reference point.

The majority of objects in our sample show pronounced Na I D absorption due to the cold neutral interstellar medium of the galaxy. To determine the kinematics of this absorption relative to the stellar component of the galaxy, we follow the fitting procedure similar to that suggested by \citet{chen10} and \citet{rupkeetal2002}. One complication is that there is a He I v5877.25\AA\ emission line very close to the relevant wavelength range. It is not a concern for \citet{chen10} who pre-subtract weak emission features, whereas we have to perform a simultaneous He I emission / Na I D absorption fit. Emission $I_{\lambda}(\lambda)$ is described by four parameters: the local continuum and the amplitude, the velocity offset and the velocity dispersion of a Gaussian He I feature. Absorption is described by another four parameters: the covering factor $C_f$, the peak optical depth $\tau_0$ of the longer wavelength component of the Na D doublet ($\lambda_R=$v5897.56\AA), the velocity dispersion of the absorbing gas $\sigma_{\rm Na}$ and its systemic velocity offset $v_{\rm Na}$. The peak optical depth of the shorter wavelength component ($\lambda_B=$v5891.56\AA) is set at $2\tau_0$ by the transition probabilities within the relevant levels of the Na atom. Then the model we fit to the data is
\begin{equation}
M_{\lambda}(\lambda)=I_{\lambda}(\lambda) \times \left [1-C_f+C_f \exp(-\tau_R(\lambda)-\tau_B(\lambda))\right],\label{eq_nad}
\end{equation}
where the optical depth of the `red' (longer wavelength) feature is given by
\begin{equation}
\tau_R(\lambda)=\tau_0\exp\left[-(\lambda-\lambda_R-v_{\rm Na}\lambda_R/c)^{2}/(2 (\sigma_{\rm Na}\lambda_R/c)^2)\right],
\end{equation}
with the `blue' (shorter wavelength) feature having the same kinematics but twice as high normalization. Example fits are shown in Figure \ref{fig:opticalfits}.

As discussed by \citet{chen10}, the covering factor, the peak optical depth and to a lesser extent the velocity dispersion are somewhat degenerate with each other because all three determine the apparent depth of the lines. Since the two components of the doublet are blended with each other, the velocity dispersion measurement relies on the wings of the absorption profile.  Furthermore, in a couple of objects the absorption appears to consist of two or more kinematically distinct components. We thus consider the $\sigma_{\rm Na}$ to be only crude estimates and wherever we report them, we do not correct them for the instrumental resolution of the SDSS spectra ($\sigma_{\rm SDSS}\simeq 64$ km sec$^{-1}$). To partially mitigate these problems, we construct a more robust physically motivated combination of these parameters, the estimated equivalent width:
\begin{equation}
{\rm EEW}\simeq 3\sqrt{2\pi}\sigma_{Na} \lambda_R \tau_0 C_f /c.
\end{equation}
This value is measured in \AA\ and for weak lines is approximately equal to the total equivalent width of Na D absorption (the equality is only approximate because this definition of EW uses only the first term of the Taylor expansion in equation \ref{eq_nad}). The outflow velocity is the most robust measure as long as the stellar redshift is reliable.

\begin{figure*}
\centering
\includegraphics[width=80mm]{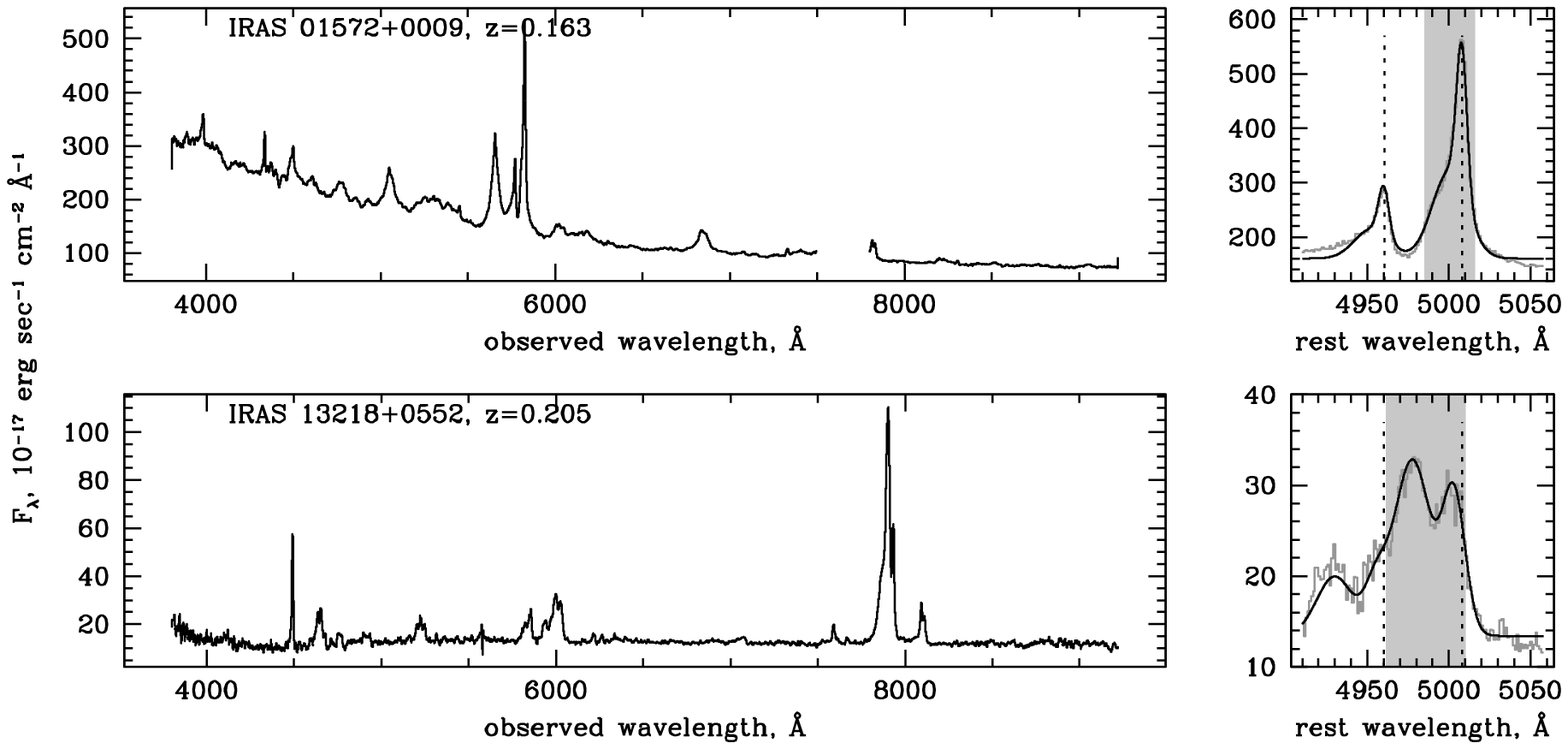}
\includegraphics[width=80mm]{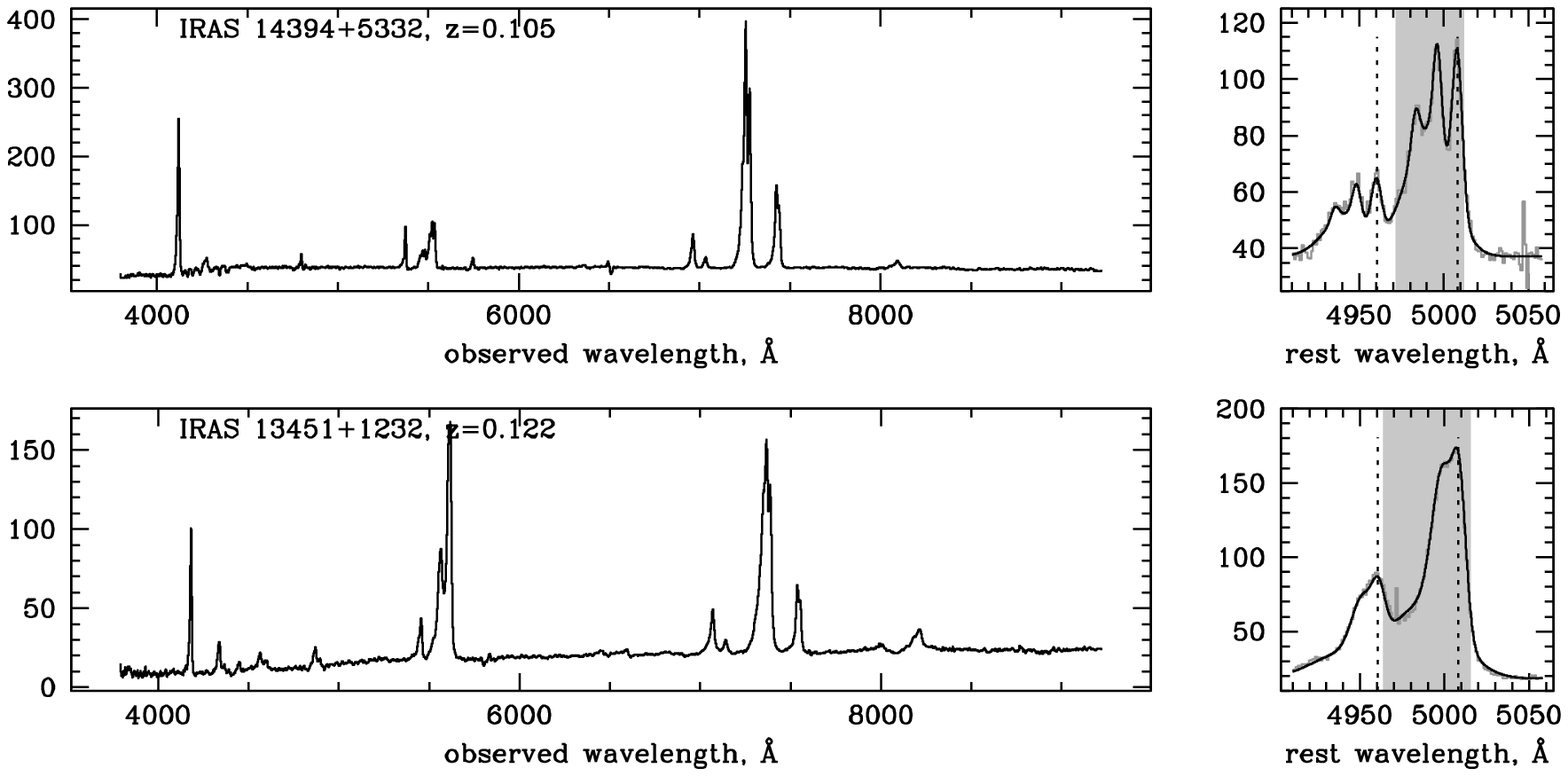}
\caption{Overall optical spectra and multi-Gaussian fits to the [OIII]$\lambda\lambda$v4960.3,v5008.2\AA\AA\ doublet in the four objects with the largest [OIII] velocity width. Dashed lines mark the laboratory wavelengths of the components of the doublet in the rest-frame of the galaxy as determined by stellar absorption features; in all four objects most of the line flux is blue-shifted relative to the expected wavelength. Shaded regions mark the range of velocities that encloses 90 per cent of the line flux, from $v_{05}$ to $v_{95}$, in each object. All four objects contain a powerful active nucleus. The alternate names and classifications from the NASA Extragalactic Database are: Mrk 1014 -- quasar (for IRAS 01572+0009; data are missing between 7500 and 7800 \AA), quasar (for IRAS 13218+0552), Seyfert 2 galaxy (for IRAS 14394+5332), 4C +12.50 -- Seyfert 1 galaxy (for IRAS 13451+1232).}
\label{fig:opticalfits2}
\end{figure*}

We now proceed to the measurements of the kinematics of the ionized gas, where we focus on the \oiii\ emission. We assume that both components of the \oiii\ doublet have the same kinematic structure and that the ratio of the peak flux density of the v4960.3\AA\ line is fixed at 0.337 relative to the stronger line at v5008.2\AA. We use one to four Gaussian components to produce a good representation of the shape of the line. As these fits are not necessarily unique, we do not attempt to interpret individual components within the line profiles. Rather, we use the fits to obtain non-parametric measures of the kinematics \citep{whit85, liu13b}. We calculate the cumulative flux as a function of velocity:
\begin{equation}
\Phi(v)\equiv\int_{-\infty}^vF_v(v'){\rm d}v',
\end{equation}
so that the total line flux is $\Phi(\infty)$ and the median velocity is the value at which 50\% of line flux accumulates: $\Phi(v_{50})=0.5 \Phi(\infty)$. Similarly, we can define velocity at 2 per cent, 5 per cent, and 95 per cent cumulative power ($v_{02}$, $v_{05}$, $v_{95}$, etc.). Furthermore, we use the velocity width containing 90 per cent of the line power, $w_{90}\equiv v_{95}-v_{05}$, measured in km sec$^{-1}$. For a Gaussian profile, this measure would correspond to 1.4$\times$ the full width at half maximum. Example fits and non-parametric measures are shown in Figure \ref{fig:opticalfits2}.

Depending on the redshift and specific dataset, the objects in this study may appear point-like or well-resolved on the sky. The datasets all have different apertures which may or may not cover the entire source. In order to avoid any aperture mismatches, we avoid using absolute fluxes from any dataset. In particular, we use line ratios derived from two lines measured from the same dataset (e.g., \molh/PAH ratios, in which both the numerator and the denominator are from \spi\ spectra), or equivalent widths, or kinematic measures -- in other words, only values that are insensitive to the exact aperture of the observations. Furthermore, in constructing line ratios we try to normalize line fluxes by star formation indicators by placing H$\alpha$, H$\beta$, Pa $\alpha$ and PAH fluxes in the denominators whenever possible.

\section{Relationships between warm \molh\ and global properties of ULIRGs}
\label{sec:morph}

Over a wide range of star formation rates, normal galaxies display a tight relationship between the luminosities of \molh\ rotational lines and the luminosities of the PAH features, with only about 0.2 dex variance if the \molh/PAH ratios \citep{rous07}. Since PAH-emitting particles are normally heated by the ultra-violet photons produced by young luminous stars, the strong correlation between PAH and \molh\ fluxes indicates that \molh\ emission in these objects is also tied to the emission of young stars.

However, this relationship breaks down at the highest galaxy luminosity. The median \molh/PAH ratios in ULIRGs exceed those seen in normal star-forming galaxies by a factor of 3 or more \citep{zaka10}, and ULIRGs no longer show a tight correlation between \molh\ and the star-formation indicators. For example, $\log({\rm S(3)}/{\rm PAH}[11.3\micron])=-1.99\pm 0.25$ for normal star-forming galaxies in the sample of \citet{smit07}, but the same ratio is $=-1.51\pm 0.31$ in our sample of ULIRGs. Similarly, in low-luminosity galaxies $\log({\rm S(1)}/{\rm PAH}[7.7\micron])=-2.48\pm 0.14$, and the same ratio is $-1.91 \pm 0.36$ among ULIRGs.

Furthermore, \molh\ emission appears to be less affected by dust absorption than PAH emission and therefore \molh\ emission likely has a more extended spatial distribution by comparison to the bulk of the star formation \citep{higd06, zaka10}. Using near-infrared ro-vibrational lines, \citet{davi03} also find hints that the ratios of \molh\ to other lines increase toward the outer parts of the ULIRGs they studied (although in this case, the overall fluxes of \molh\ lines are not inconsistent with excitation by ultra-violet photons).

In this section and the next one, we investigate the observed relationships between the amount of \molh\ emission and other properties of ULIRGs in order to determine the origin of this excess emission. In most cases we use the \molh/PAH ratios to quantify the excess of \molh\ over that expected the star formation rates.

\subsection{Effects of an active nucleus}
\label{sec:AGN}

We first investigate whether the presence of an AGN affects the \molh/PAH ratios. As the first step, following \citet{yuan10}, we split the sample into four subsamples by optical classification: purely star-forming objects whose spectra are consistent with those of HII regions (13 objects), AGNs (37 objects, combining type 1 and type 2 objects), and star formation / AGN composites (53 objects). We use the KS test to determine whether any two of these subsamples show statistically different distributions of \molh/PAH or \texc.

\begin{figure*}
\centering
\includegraphics[width=80mm]{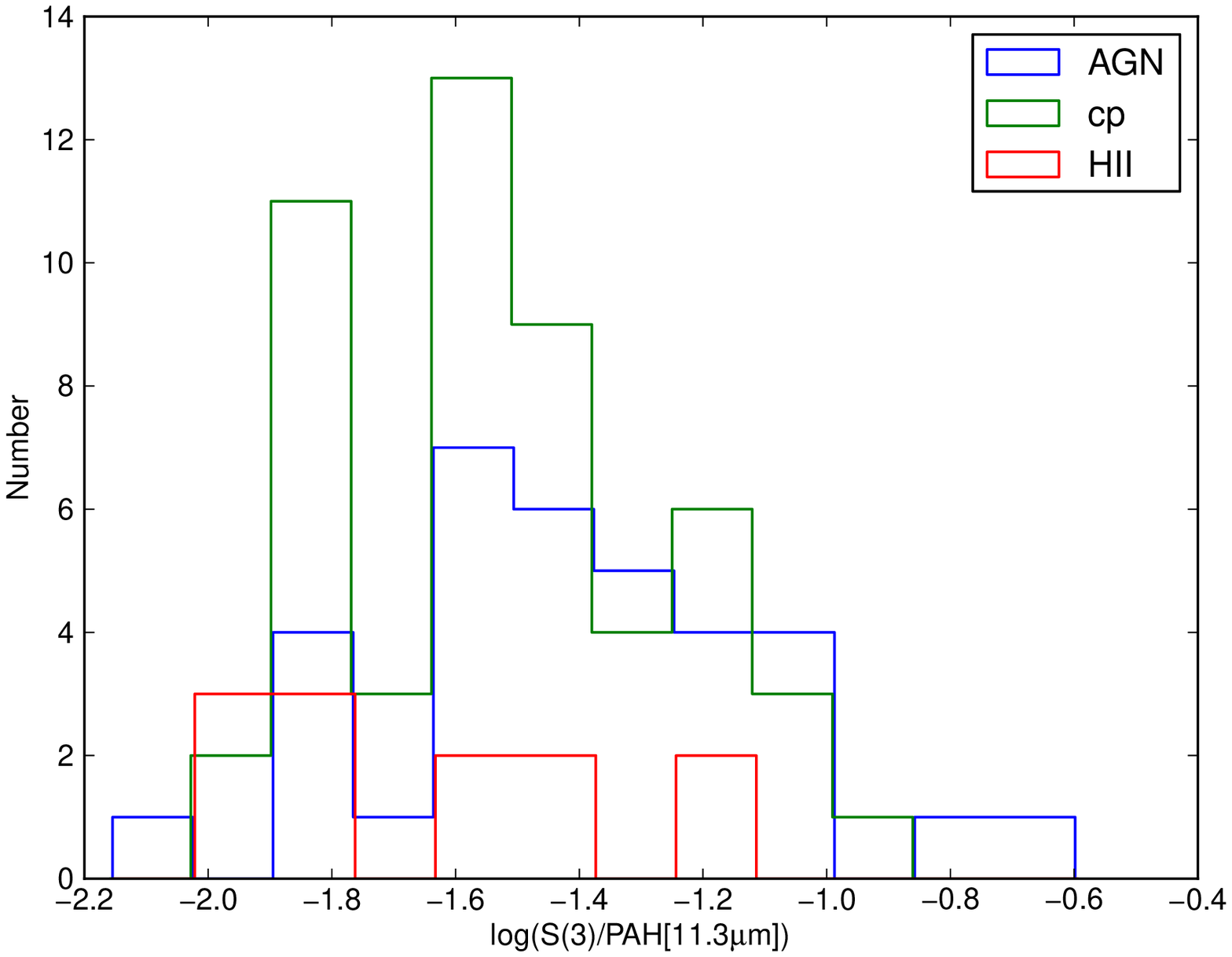}
\includegraphics[width=80mm]{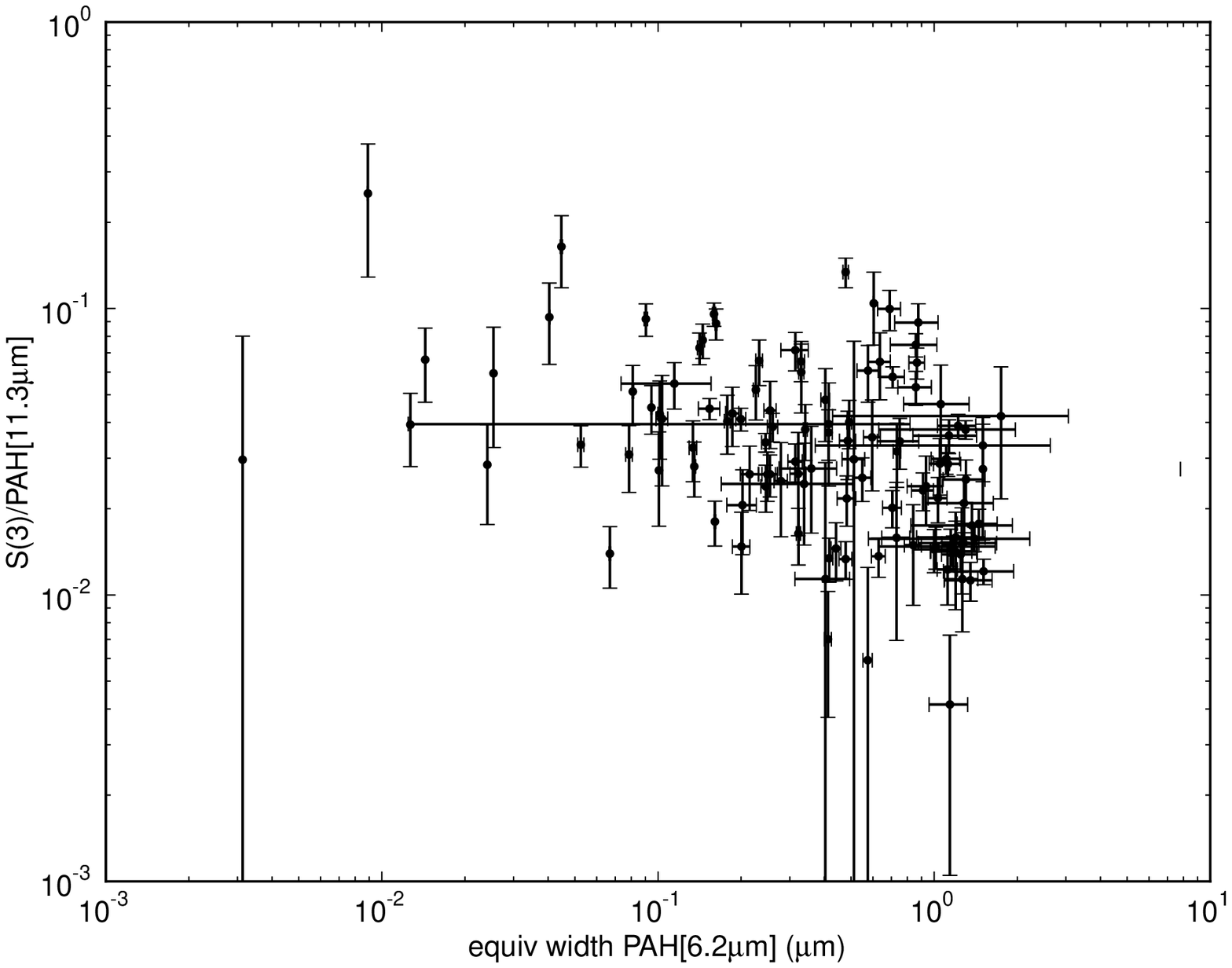}
\caption{{\bf Left:} The distributions of \molh/PAH ratios by optical classification. We see slightly higher \HPAH\ ratios in galaxies classified as AGNs or star-formation/AGN composites than in purely star-forming galaxies ($p_{\rm KS} = 0.04$ and $p_{\rm KS} = 0.064$, respectively). {\bf Right:} We find an anti-correlation between \HPAH\ and equivalent width of PAH[6.2$\mu$m] ($p_{\rm S} = 3\times 10^{-6}$) indicating that \molh\ excess is stronger in galaxies with a larger relative contribution of the active nucleus. (For comparison, the median value of \HPAH\ in low-luminosity star-forming galaxies is 0.01, below most values found in our sample of ULIGRs.) \label{fig:AGN}}
\end{figure*}

We find that galaxies classified as AGNs or composites have slightly higher \HPAH\ emission than star-forming galaxies (Figure \ref{fig:AGN}, left), which is similar to the trends seen in lower luminosity galaxies \citep{rigo02, rous07}. Star-forming galaxies have a median log(\HPAH) = $-1.83 \pm 0.32$. Composite galaxies have median log(\HPAH) = $-1.54 \pm 0.28$ ($p_{\rm KS} = 0.064$ between this population and the star-forming galaxies). Galaxies classified as AGNs have median log(\HPAH) = $-1.43 \pm 0.32$ ($p_{\rm KS} = 0.044$ between active nuclei and star-forming galaxies). Thus there is a general, albeit weak, trend that the \molh/PAH ratios increase as the fractional contribution of the active nucleus to the optical diagnostics increases. The median excitation temperature measured from the S(3)-S(1) transition is 406 K. We do not see any difference in \texc\ between galaxies of different optical classification.

Optical emission lines are not necessarily the most reliable diagnostic of the presence of an active nucleus in objects that are heavily obscured by dust. Furthermore, because of the sensitivity of optical diagnostics to the spatial distribution of dust extinction they cannot be used to determine the relative contributions of the star formation and of the accretion onto a supermassive black hole to the bolometric luminosity of ULIRGs. Mid-infrared diagnostics are less affected by obscuration and thus may be more reliable. In particular, the equivalent widths of the PAH features can serve as a measure of the contribution of the active nucleus to the total energy budget of a ULIRG \citep{iman00, armu07, spoo07}. Indeed, the luminosity of the PAH features is a good indicator of star-formation rates \citep{rous01, dale02}, whereas the continuum is produced by dust heated both by star formation and the active nucleus. Thus, low equivalent widths of PAH emission are expected in objects powered predominantly by black hole accretion. The equivalent width of the 6.2\micron\ PAH feature is a particularly good indicator. The active nucleus, if any, makes a two-fold contribution: it dominates the continuum at these short wavelengths (and correspondingly warmer dust temperatures) and it suppresses the emission of the feature itself \citep{diam10}.

We find an anti-correlation between \HPAH\ and the equivalent width of the PAH[6.2\micron] feature ($p_{\rm S} = 3\times10^{-6}$; Figure \ref{fig:AGN}, right). Thus the mid-infrared data indicate that the presence of an active nucleus is either associated with the destruction of the PAH[6.2$\mu$m] emission, as was suggested by \citet{smit07} and \citet{diam10}, or with production of excess \molh, or both. We do not see any relationship between \texc\ and PAH[6.2$\mu$m] equivalent width. 

There are some drawbacks to using equivalent widths of PAHs as the measure of the AGN contribution. For example, low equivalent width of PAHs can be due to low metallicity rather than high AGN contribution, and the lack of excitation of PAHs by emission from active nuclei is not well understood (\citealt{farr07} provide a detailed discussion). Furthermore, in Figure \ref{fig:AGN} PAH luminosities (albeit of two different features) appear both in the numerator of the horizontal axis and the denominator of the vertical axis, which could potentially give rise to a spurious anti-correlation. Thus, we examine the relationships between \HPAH\ and other measures of AGN activity suggested in the literature, especially ones based on high-ionization emission lines \citep{farr07}. In particular, we find that \HPAH\ positively correlates with \nev/\neii\ ($p_{\rm S}=10^{-4}$), confirming our finding that excess \molh\ is more pronounced in objects with stronger AGN activity. Objects with higher \HPAH\ tend to have higher \oiv/\neii\ ratios as well, but the correlation is marginal at best ($p_{\rm S}=0.1$). The main drawback to using these diagnostics is that \nev\ and \oiv\ are reliably measured in only 20-30 of our sources. 

To summarize, we use a variety of optical and infrared diagnostics to test the relationship between the excess \molh\ and the presence of AGN activity. While none of the diagnostics provides a perfect measure of the AGN contribution to the overall energetics, in all cases we see a trend of higher \molh/PAH ratios in objects with stronger AGN activity. 

\subsection{Effects of galaxy morphology and merger stage}

Numerical simulations have shown that major galaxy mergers are able to induce nuclear gas flows and produce strong starburst episodes \citep{barn91, barn96, miho96, miho94}. The nuclear gas inflows during the merger sequence are due to the strong non-axisymmetric perturbations in the gravitational potential induced by each galaxy on its companion. While the gas may have started in rotational equilibrium within each galaxy disk, the non-axisymmetric potential induced during the merger results in the appearance of radial orbits which do not conserve angular momentum and funnel the gas toward the centre of the galaxy \citep{binn08}, where it compresses, shocks and gives rise to new stars. The same non-axisymmetric potential that promotes the driving of gas into the nuclear region of the galaxy also promotes the development of long tidal tails \citep{barn91}.

Tracing of these phenomena throughout the merger \citep{nara10b, nara10a, torr12} gives insight into the timing of these events. After the first close passage, nuclear mass inflow and the resulting star formation rate increase steadily for $\sim$100 Myr and remain fairly high while the galaxies conduct their quasi-Keplerian orbital motion around each other, moving out to the apo-galacticon (which can be many tens of kpc) and coming back for the second close approach.

Because gas-rich galaxy mergers are dissipative, the second close approach tends to be closer than the first one, and subsequently the galaxies separate by only a few kpc for only a short period of time before the final coalescence. The typical time scale for the entire merger from the first approach to the final coalescence is on the order of 1 Gyr, with 90 per cent of it taken by the time between the first and the second close passage \citep{torr12}. The gas inflow rate and the star formation rate peak strongly between the second close approach and the final coalescence, but for a relatively short period (0.1 Gyr; \citealt{nara10b, nara10a, torr12}).

\begin{figure*}
\centering
\includegraphics[width=80mm]{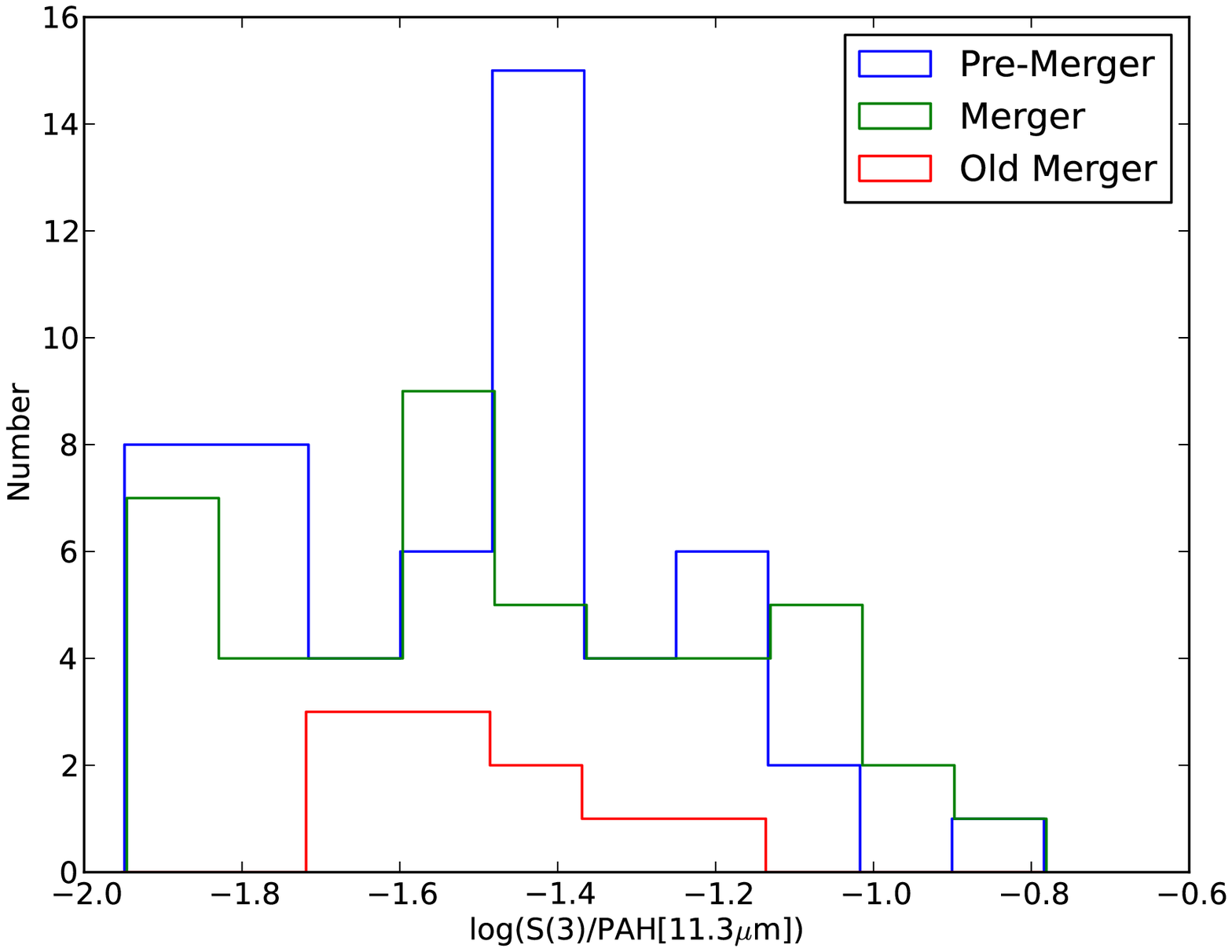}
\includegraphics[width=80mm]{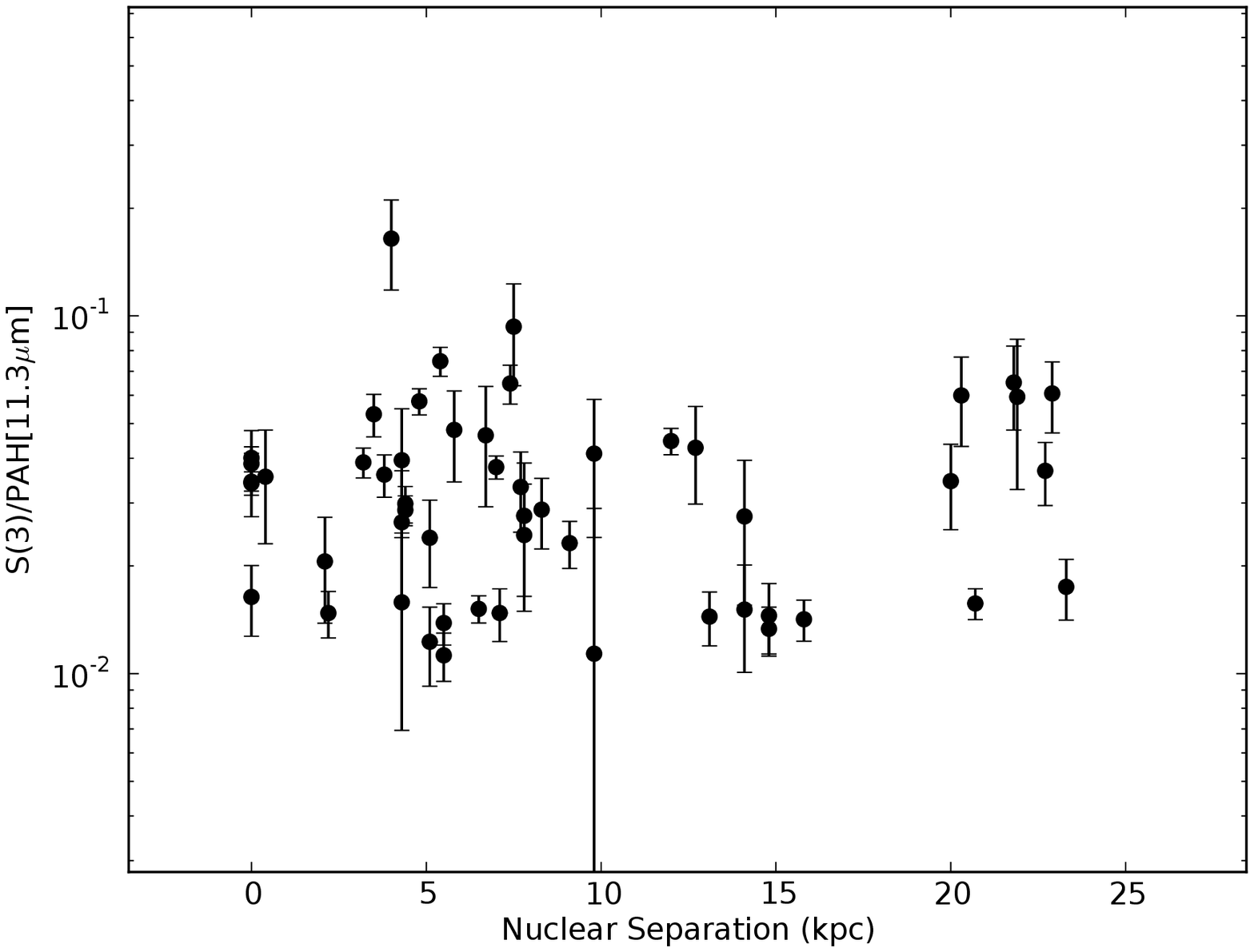}
\caption{{\bf Left:} S(3)/PAH[11.3$\mu$m] is constant across merger class ($p_{\rm KS}=0.2-0.4$ between any pair of categories). {\bf Right:} S(3)/PAH[11.3$\mu$m] is uncorrelated with projected nuclear separation ($p_{\rm S}>0.9$).\label{fig:mergclass_ns}}
\end{figure*}

ULIRGs are defined by their high infrared luminosity, which is largely due to a very high star formation rate, and thus it is not surprising to find the vast majority of ULIRGs in merging systems \citep{sand96}. Given the importance of the merger process in the development of ULIRGs, we look for possible connections between \molh\ and morphological features associated with merger activity. We do not find any differences in the distribution of \molh/PAH between the Pre-Merger, Merger and Old Merger classes ($p_{\rm KS}=0.2-0.4$; Figure \ref{fig:mergclass_ns}) nor do we see a difference in the distribution of \texc\ between any merger classes ($p_{\rm KS}=0.3-0.9$). Thus we do not confirm the trend seen by \citet{petr10} who find higher excitation temperatures of \molh\ in advanced mergers within a somewhat less luminous sample of galaxies \citep{petr11}.

Within the Pre-Merger class, the easily measurable apparent nuclear separation does not yield a direct measure of merger progress: the approaching galaxies on first approach, the receding galaxies right past the first approach on the way to apo-galacticon and the approaching galaxies on the way to the second approach could all have the same nuclear separation, although after the second approach the merging pairs do not tend to separate much \citep{torr12}. Furthermore, projection effects make it difficult to distinguish widely separated galaxies which happen to be close to the line of sight from pairs at smaller physical separations. Nevertheless, in a large unbiased survey of galaxy pairs the apparent nuclear separation correlates with the excess star formation rate induced during the interaction \citep{elli13}. Thus the apparent nuclear separation must provide some measure of the interaction stage.

We do not find any correlation between \HPAH\ and projected nuclear separation ($p_{\rm S}=0.99$) nor between excitation temperature \texc\ and projected nuclear separation ($p_{\rm S}=0.95$). We previously hypothesized that shocks produced directly in the galaxy collisions could be responsible for the excess \molh\ observed in ULIRGs \citep{zaka10}, by analogy to other systems in which high-velocity intra-group collisions are associated with warm \molh\ emission \citep{appl06, pete12, cluv13}; our new results indicate that this hypothesis is an unlikely explanation. Indeed, one would expect to see the strongest effect of galaxy collision shocks in pairs with smallest nuclear separations or in the Merger class.

\begin{figure*}
\centering
\includegraphics[width=80mm]{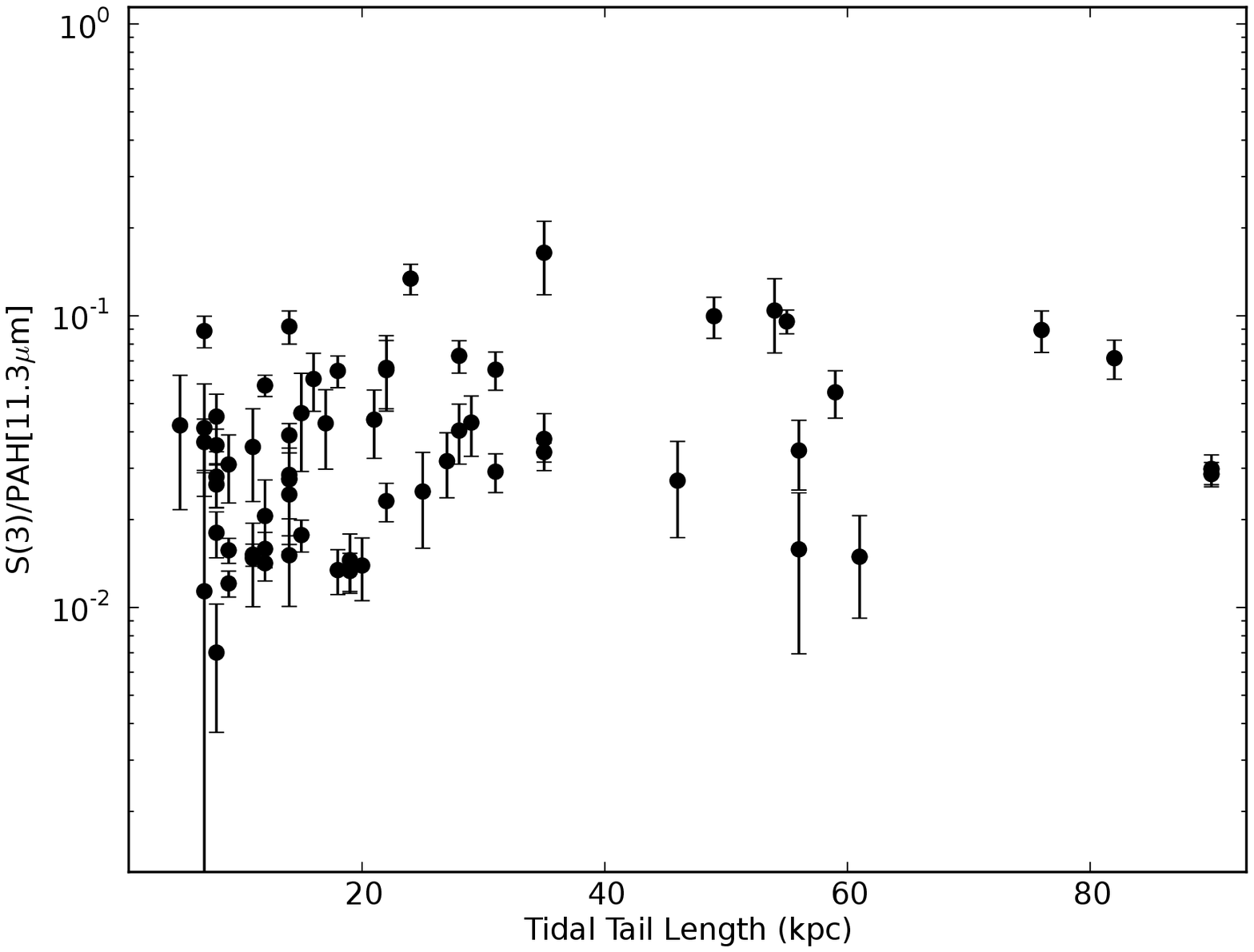}
\includegraphics[width=80mm]{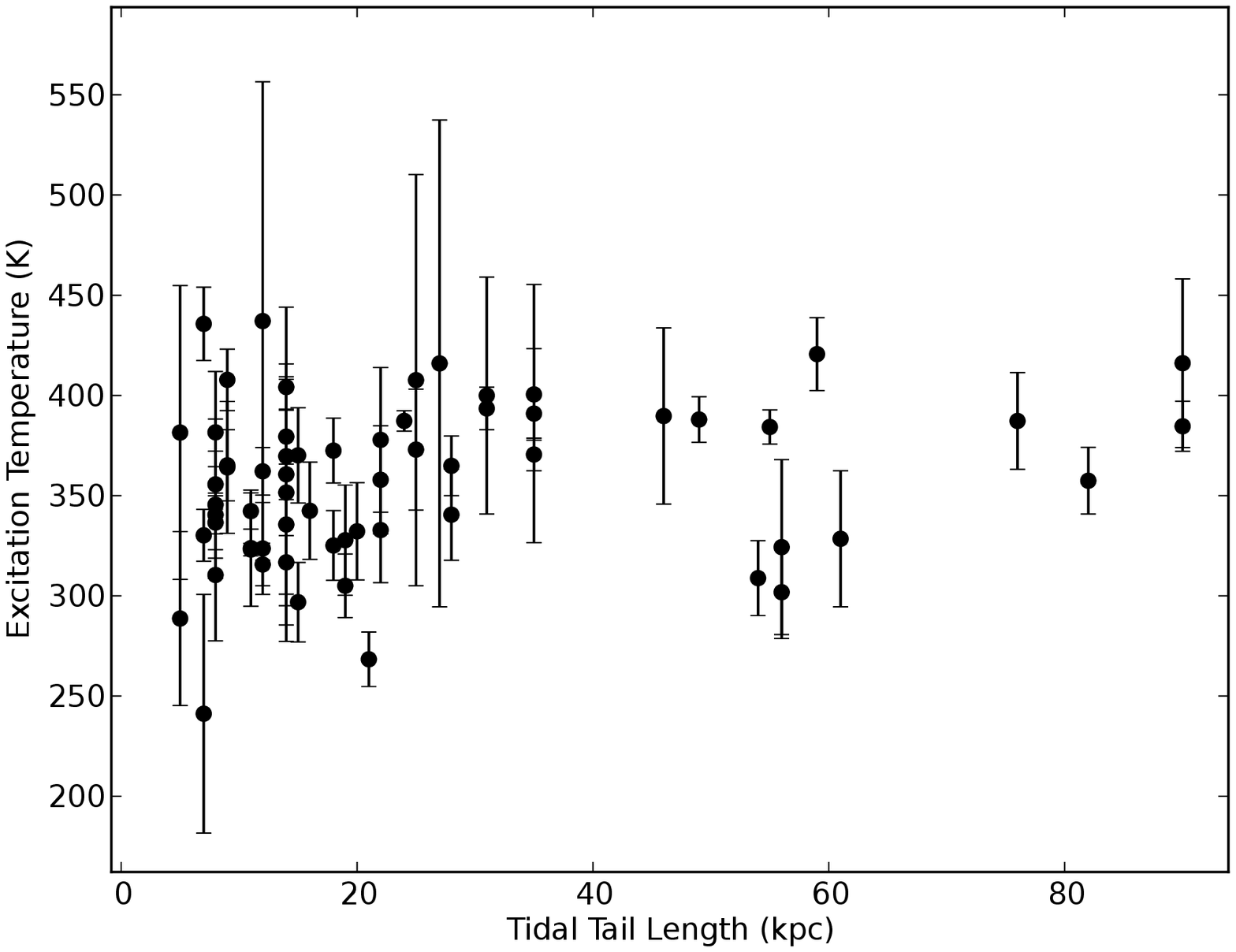}
\caption{The length of tidal tails correlates weakly with \HPAH\ (left, $p_{\rm S}$ = 0.01) as well as with \texc\ (right, $p_{\rm S}$ = 0.04).\label{fig:tidal_tails}}
\end{figure*}

Tidal tails are another useful feature in determining the stage of the merger. The production of tidal tails is dependent on the collision geometry with the longest tidal tails developing in collisions where there is a resonance in rotational motion in prograde encounters \citep{toom72}. In general, tidal tail length monotonically increases as a function of time, although again the projection effects and the disappearance of features as they fall below the surface brightness detection threshold make it difficult to use them as a precise timer of the merger. We do find an association between the length of the tidal tails in our objects and excess \molh\ emission ($p_{\rm S} = 0.01$) as well as a more tentative association with \texc\ ($p_{\rm S} = 0.04$; Figure \ref{fig:tidal_tails}). Pure star formation galaxies tend to have tidal tails less frequently (2 of 13, 15 per cent) compared to composites (34 of 53, 64 per cent) and active nuclei (24 of 37, 65 per cent). However, there is no correlation between tidal tail length and the equivalent width of PAH[6.2$\mu$m] -- a measure of the AGN contribution.

For the 36 objects with the SDSS spectra, we estimate the age of the starburst by using the strength of the 4000\AA\ break \citep{kauf03a} which is parametrized by the ratio of average flux densities on either side of the break,  $D_n(4000)=\langle f_{\nu}(4000-4100{\rm \AA}) \rangle/\langle f_{\nu}(3850-3950{\rm \AA})\rangle$. As a stellar population ages and becomes redder, the value $D_n(4000)$ monotonically increases; the precise calibration of this parameter with starburst age is presented by \citet{kauf03a}. Before this value can be measured, the emission lines need to be subtracted to reveal the pure stellar continuum of the galaxy. In our objects, a particularly strong line in the relevant wavelength range is [Ne {\sc iii}]$\lambda$v3870\AA, which we either interpolate over using surrounding continuum or exclude from the calculation of $\langle f_{\nu}(3850-3950{\rm \AA}) \rangle$ entirely -- the results are insensitive to the specific procedure used.

We find that $D_n(4000)$ values range between 1.0 and 1.6, corresponding to starburst ages between $\tau_{\rm SB}=10$ Myr and 1 Gyr. The median and the standard deviation of $D_n(4000)$ is 1.20$\pm$0.14, corresponding to the median age $\tau_{SB}=250$ Myr. We find no correlation between \HPAH\ and the age of the starburst ($p_{\rm S}>0.08$), and within the range of starbust ages in our sample there is no particular value associated with stronger \molh\ emission.

To summarize, objects with strongest excess \molh\ emission do not have preference for a specific merger stage or starburst age. However, the correlation with the tidal tail length indicates that the amount of excess \molh\ emission and the excitation temperature increase with time during galaxies' orbit after the first close approach. 

\section{Relationship between warm \molh\ and other gas phases}
\label{sec:phases}

The interstellar medium of a normal galaxy is comprised of gas at a wide range of densities and temperatures. Each phase of this medium produces its own emission and absorption depending on the dominant radiative processes in this phase. In this section, we use the rotational lines of \molh\ to probe the warm molecular gas phase of the interstellar medium; optical, near-infrared and mid-infrared recombination emission lines to probe the ionized gas phase; and the Na D doublet to probe the neutral gas phase. Using these diagnostics, we investigate the relationships between the kinematics and the physical conditions of gas in these three phases.

\subsection{Optical lines of ionized and neutral gas}

\begin{figure}
\centering
\includegraphics[width=80mm]{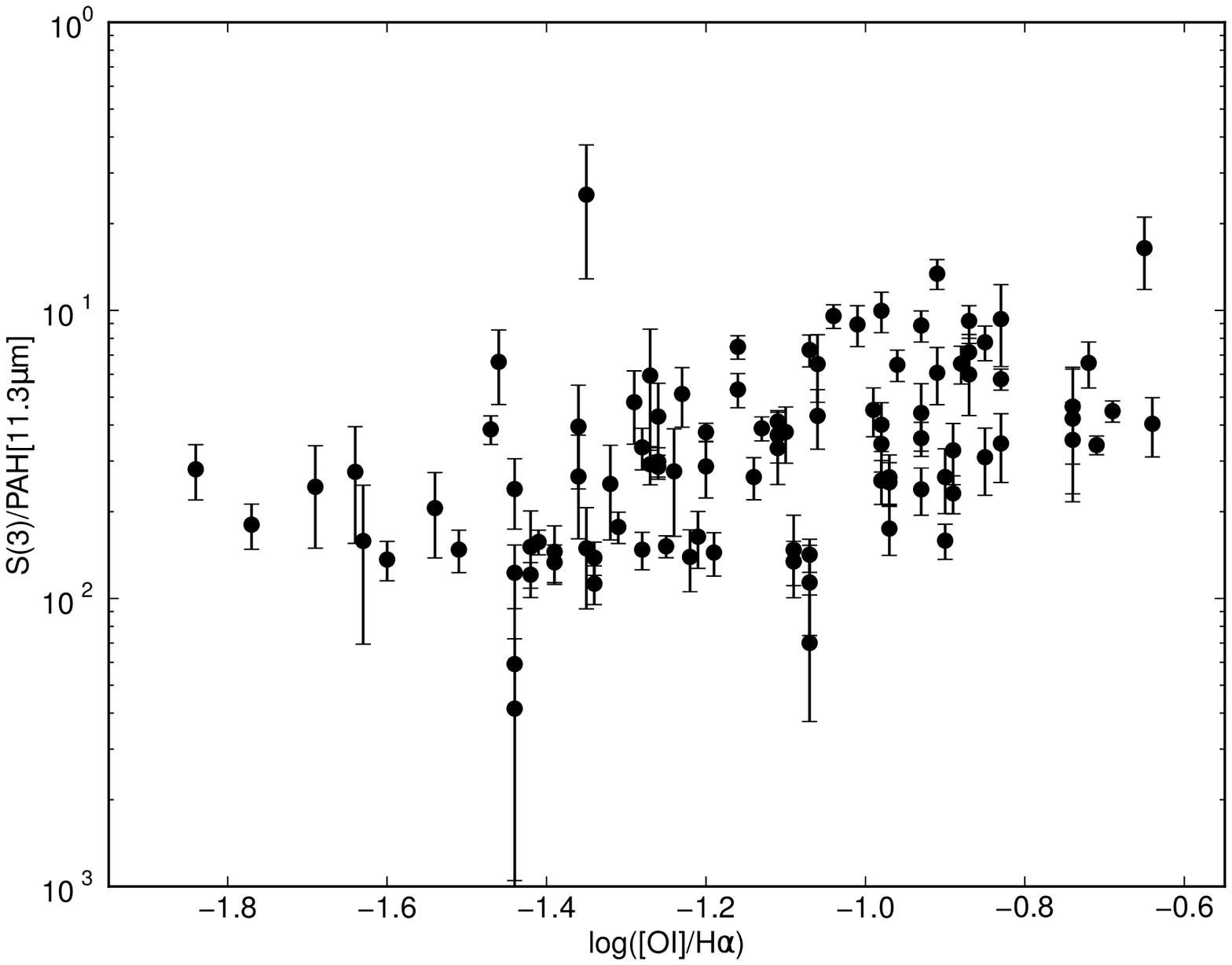}
\caption{We look at the relationships between \molh/PAH and optical line ratios, such as \oir, \niir, \siir, and \oiii/H$\beta$. The strongest correlation is between S(3)/PAH[11.3$\mu$m] and \oir\ ($p_{\rm S}=10^{-8}$) shown here. \label{linefig}}
\end{figure}

Optical emission lines produced by ionized and neutral gas carry information about the physical conditions in the gas and its source of ionization and excitation. We start by looking at the correlations between \molh/PAH and the optical line ratios, such as \niir, \oir, \siir\ and \oiii/H$\beta$. The combination of these measurements can be used to determine gas metallicity, ionization parameter, and relative strengths of the contributions of shock- and photo-ionization. We find a positive correlation between \HPAH\ and [OI]/H$\alpha$ ($p_{\rm S}=10^{-8}$, Figure \ref{linefig}), but weaker, if any, correlations between \HPAH\ and the other ratios: $p_{\rm S}=0.03$ for \niir, $p_{\rm S}=10^{-3}$ for \siir, and $p_{\rm S}=2\times 10^{-3}$ for \oiii/H$\beta$.

The relative strengths of these correlations -- in particular, the fact that the strongest correlation is with \oir\ -- suggest that the \molh/PAH ratios are driven in large part by the relative contribution of shocks. Indeed, while all these four optical line ratios increase as the fractional contribution of shocks increases, the \oir\ ratio has the strongest sensitivity on the shock fraction. As shown by \citet{rich11}, at a fixed ionization parameter, increasing the shock contribution to gas ionization from 5 per cent to 80 per cent increases \oiii/H$\beta$ by $\sim$0.4 dex, \siir\ and \niir\ by $\sim$0.8 dex, and \oir\ by $\sim$1.2 dex. The values of \oir\ that we see in our sample span the entire range of models probed by these authors.

Conversely, at a fixed shock fraction the \oir\ line ratio is the least sensitive to the variations in the ionization parameter. If \molh/PAH variations among the objects in our sample were driven by the variations in the ionization parameter, we would expect a weaker, if any, correlation with \oir\ than with the other line ratios. Thus, although AGNs have somewhat higher \molh/PAH ratios (as we saw in Sec. \ref{sec:AGN}), the mere presence of an AGN or even its fractional contribution to the ionization balance do not appear to be the primary driving factors for the amount of \molh\ emission. This conclusion is similar to that reached by \citet{rous07} who considered excess rotational \molh\ emission in lower-luminosity galaxies. These authors also found that direct excitation by the emission from an AGN (specifically, by the X-rays from the AGN) is insufficient to produce the observed excess of \molh.

Applying extinction corrections derived from Balmer decrements results in negligibly small changes to all ratios (purposefully chosen to use features that are close together in wavelength) except \oir. A correlation between extinction-corrected \oir\ and \molh/PAH is present with $p_{\rm S}=10^{-5}$ and is qualitatively similar to that shown in Figure \ref{linefig}. Depending on the spatial extents of \oi-emitting shocks relative to the bulk of the obscuration occurring in the dusty medium of ULIRGs, the extinction correction of the shock diagnostics may or may not be warranted. If for example a compact starburst (where most of ionized hydrogen emission originates) is responsible for shocks driven in a more extended quasi-neutral medium, then the Balmer decrements yield an overestimate of actual extinction, and the correlation seen in Figure \ref{linefig} is then a more accurate representation of the intrinsic conditions. 

\begin{figure*}
\centering
\includegraphics[width=80mm]{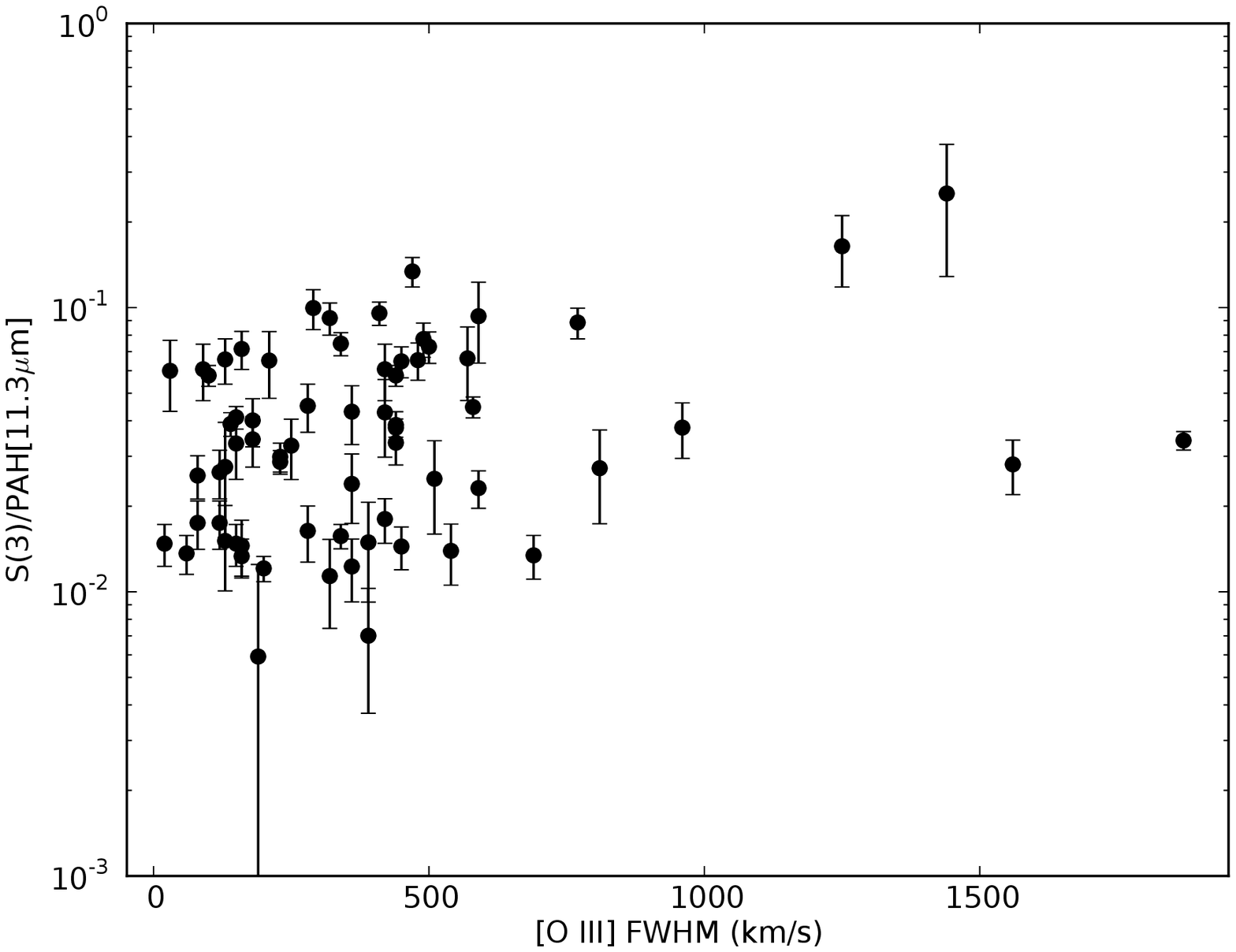}
\includegraphics[width=80mm]{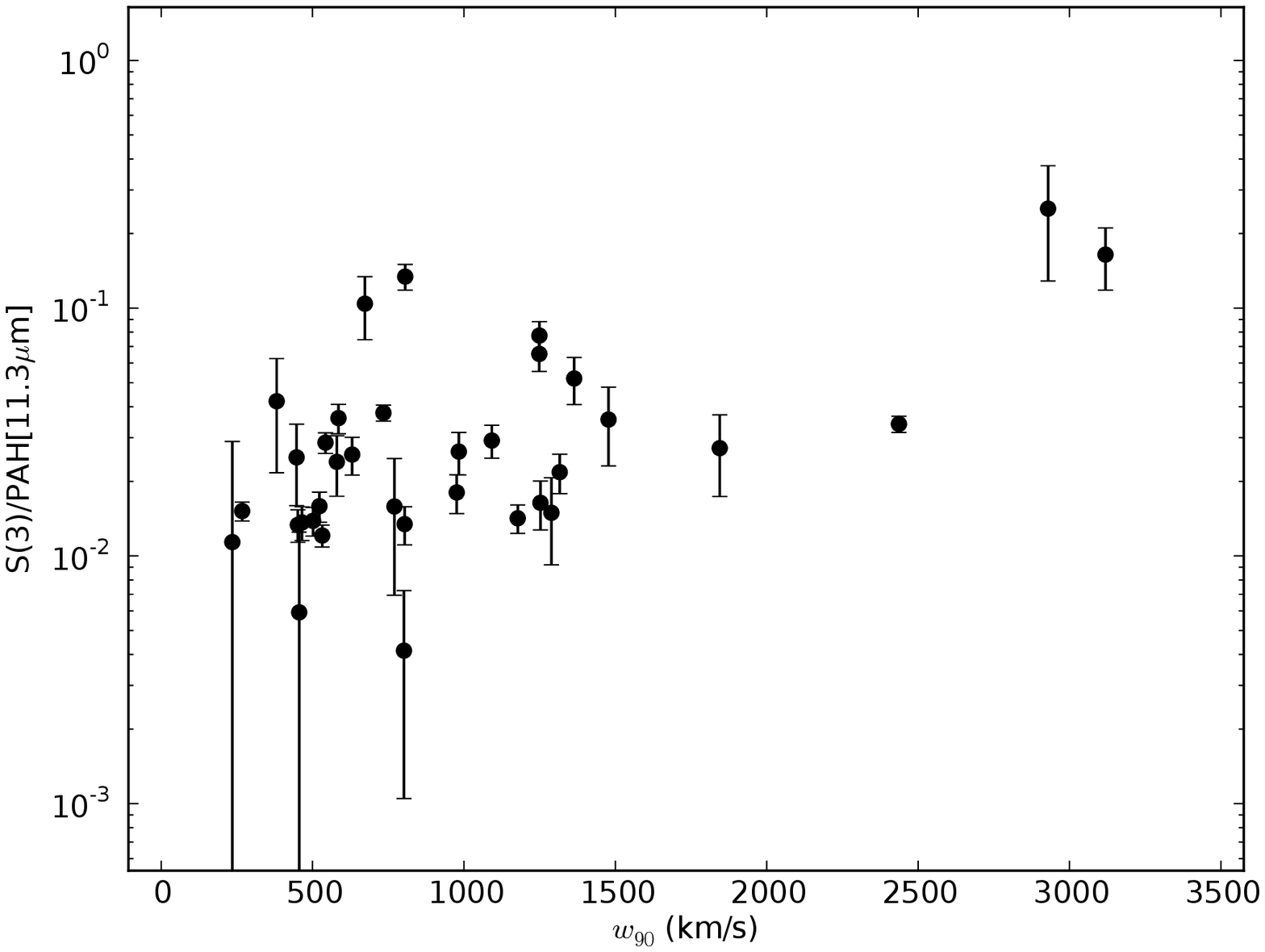}
\caption{{\bf Left:} The relationship between \HPAH\ and the full width at half maximim of \oiii\ taken from \citet{veil99} for the full sample shows a tentative positive correlation ($p_{\rm S}=0.03$). {\bf Right:} The relationship between \HPAH\ and the velocity widths comprising 90 per cent flux $w_{90}$ for the subsample of 36 objects with SDSS spectra shows a positive correlation ($p_{\rm S}=0.002$). \label{fig:oiiiwidth}}
\end{figure*}

Relationships between the kinematics of the ionized gas and the \HPAH\ ratios offer further clues regarding the origin of the \molh\ emission in ULIRGs. We find a positive correlation between \HPAH\ and the velocity widths of the optical emission lines (Figure \ref{fig:oiiiwidth}). The correlation is present both when we consider the single-Gaussian fits to low-resolution optical spectroscopic data and when we perform the multi-Gaussian fitting more sensitive to weak broad components in emission lines. There is no correlation between the excitation temperature of \molh\ and optical line widths. 

Like many other authors (e.g., \citealt{soto12}), we find that the optical emission line gas shows predominantly blue-shifted asymmetries (e.g., Figure \ref{fig:opticalfits2}). Taken by itself, this observation indicates that the ionized gas in ULIRGs is likely embedded in an outflow, which may be intrinsically symmetric, but its redshifted part (being located on the other side of the galaxy from the observer) is dimmed by the intervening galaxy material. Furthermore, in ULIRGs there is a strong correlation between the line-of-sight velocity dispersion of the ionized gas and the shock contribution to its ionization balance \citep{rich11, soto12}. This likely means that higher outflow velocities (reflected in the line-of-sight velocity dispersion) produce stronger shocks which are more likely to ionize the clouds they impact.

As already mentioned in Sec. \ref{sec:morph}, we do not see any correlations between \molh/PAH and the ages of the stellar populations (as measured by $D_n(4000)$); neither do we find one between the velocity of the outflow (as measured either by $v_{02}$ or by $w_{90}$) and the ages of the stellar populations. Thus there is no particular age of the stellar populations when the outflows have highest velocity. This is likely a result of the long duration of the starburst induced on the initial approach. Because gas is continually supplied throughout the merging process, the $D_n(4000)$ is unlikely to be a particularly accurate clock of the merger.

The relationships between excess \molh\ emission, optical line ratios and optical line kinematics suggest a picture in which a galaxy-wide outflow that shocks and entrains the interstellar medium is responsible for some of the optical line emission as well as much of the warm \molh\ emission. Stronger shocks appear to be associated with higher outflow velocities and stronger \molh\ emission. Unfortunately because of the low resolution of the \spi\ spectroscopic data we cannot establish whether the ionized gas outflows and the molecular gas outflows share the same kinematics. \citet{rupk13b} demonstrated that it is not necessarily the case: in a heavily obscured ULIRG which likely just recently started driving an outflow, the warm molecular hydrogen has significantly slower outflow velocities than those seen in the ionized hydrogen. It is tempting to conclude that the higher density molecular outflow with its greater inertia takes longer to accelerate than the ionized gas component, although a larger sample size would be desirable to test this hypothesis. Furthermore, it would be interesting to look closer at the kinematics of the quasi-neutral gas which we have demonstrated may be more strongly coupled to the molecular outflow than the ionized gas probed by the standard diagnostics like \oiii\ in the optical and Pa $\alpha$ in the near-infrared. 

What is the origin of these outflows? Most of them are likely driven by explosions of supernovae associated with the recent star formation activity \citep{soto12}. But while the majority of the ULIRGs are likely powered by high rates of star formation, the four objects with the largest \oiii\ velocity widths in our sample (Figure \ref{fig:opticalfits2}) are all optically classified as AGNs, one of them radio-loud. How do AGNs affect the excitation of \molh? \citet{rous07} find that X-rays are in general insufficient to explain the excess \molh/PAH ratios in nearby low-luminosity galaxies. Therefore, given the correlation between \molh\ and \oiii\ emission line width in our sample, we propose that \molh\ emission in all ULIRGs is dominated by shocks, both in starburst-dominated objects (where the winds are driven by supernovae explosions) and in the AGN-dominated ones \citep{diam10}. In the latter case, the outflow may be driven either by the relativistic jets or by the radiation pressure due to the accretion onto the supermassive black hole \citep{murr95, prog00}.

\begin{figure}
\centering
\includegraphics[width=80mm]{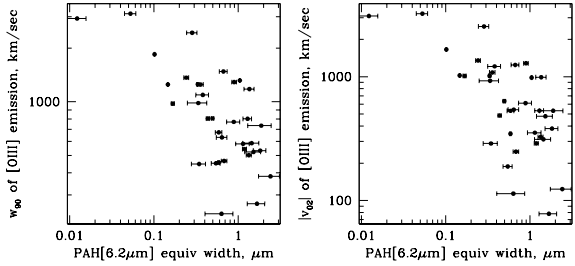}
\caption{Two different measures of wind velocity as a function of the contribution of the active nucleus to the bolometric luminosity strongly correlate with the equivalent with of PAH[6.2\micron] ($p_{\rm S}=2\times 10^{-4}$ and $6\times 10^{-5}$) which serves as an indicator of the AGN contribution to the bolometric luminosity. An object powered by an AGN and by star formation in roughly equal measures has EW(PAH[6.2\micron]) of about 0.3\micron\ \citep{armu07}. Thus objects on the left-hand side of each panel are AGN-dominated and those on the right-hand side are star-formation-dominated. Using PAH[6.2\micron] equivalent widths from \citet{veil09a} which are corrected for water ice absorption does not change the qualitative character of the relationships, but shifts the abscissas to the left by a median of 0.16 dex. \label{fig:vel}}
\end{figure}

\citet{rupk05} and \citet{rupk13a} suggest that outflow velocity can be used as a discriminant of starburst-driven vs AGN-driven winds. Using the highest blueshifted velocity (in our notation it is the velocity at the 2 per cent of the line flux $v_{02}$), they find median values of $v_{02}\simeq -1000$ km sec$^{-1}$ for AGN-driven winds and $-500$ km sec$^{-1}$ for starburst-driven ones. We explore this dichotomy further using the 36 objects with the SDSS spectra in our sample. In Figure \ref{fig:vel}, we show the outflow velocity -- parametrized both by the line width $w_{90}$ and the maximum blueshift velocity $v_{02}$ -- as a function of the equivalent width of the PAH[6.2\micron] emission used as a measure of the AGN contribution to the bolometric luminosity.

The AGN-dominated objects on the left-hand side of each panel show significantly higher outflow velocities by both measures than the starburst-dominated objects on the right. The characteristic values we obtain for $v_{02}$ are in agreement with those found by \citet{rupk13a}. Thus it appears likely that the objects with the highest \oiii\ outflow velocities and velocity widths -- which also tend to show high \molh/PAH ratios -- have winds driven by the AGN activity. Because of the obscuration of the central AGN, these winds would not necessarily be photo-ionized \citep{soto12} with characteristic AGN-like line ratios, but would be shock-ionized just like supernova-driven winds. Aside from the buried energy source -- supernova explosions vs radiative pressure -- the physics of galaxy-wide starburst-driven and AGN-driven outflows is apparently similar.

The observed striking difference in outflow velocities between AGN-dominated and star-formation-dominated ULIRGs has the potential to become a useful diagnostic of the powering source of the outflows. Qualitatively, this velocity difference was predicted in numerical simulations of mergers by \citet{nara08}. In these simulations, the relatively high velocities of AGN-driven outflows are explained by a higher rate of energy injection into the outflow compared to the supernova-driven winds, even though the AGN-dominated phase may be comparatively shorter. Further numerical work will help determine whether the model velocity profiles of emission lines are consistent with those observed.

Our conclusion regarding the origin of \molh\ emission in AGNs is somewhat different from that reached by several other groups. \citet{mour89} argue that X-ray emission is likely the dominant source of excitation of ro-vibrational \molh\ lines, and \citet{rigo02} suggest the same for the rotational \molh\ lines, altough \citet{rous07} find X-ray emission insufficient and suggest that starburst-driven winds of the host galaxy provide the \molh\ excitation. The luminosity of the AGN appears to be the determining factor of whether it shows a large galaxy-wide outflow \citep{liu13a}, with only the most luminous obscured quasars routinely displaying one. The objects that we consider in this paper are much more luminous than those of \citet{mour89}, \citet{rigo02} and \citet{rous07} -- indeed, if the AGN dominates the bolometric luminosity of a ULIRG then it is comparable to the objects studied in \citealt{liu13a}. This significant step up in luminosity may explain why we see the signatures of AGN-driven winds not apparent in the previous samples.

\subsection{[Fe II] emission}

\begin{figure*}
\centering
\includegraphics[width=80mm]{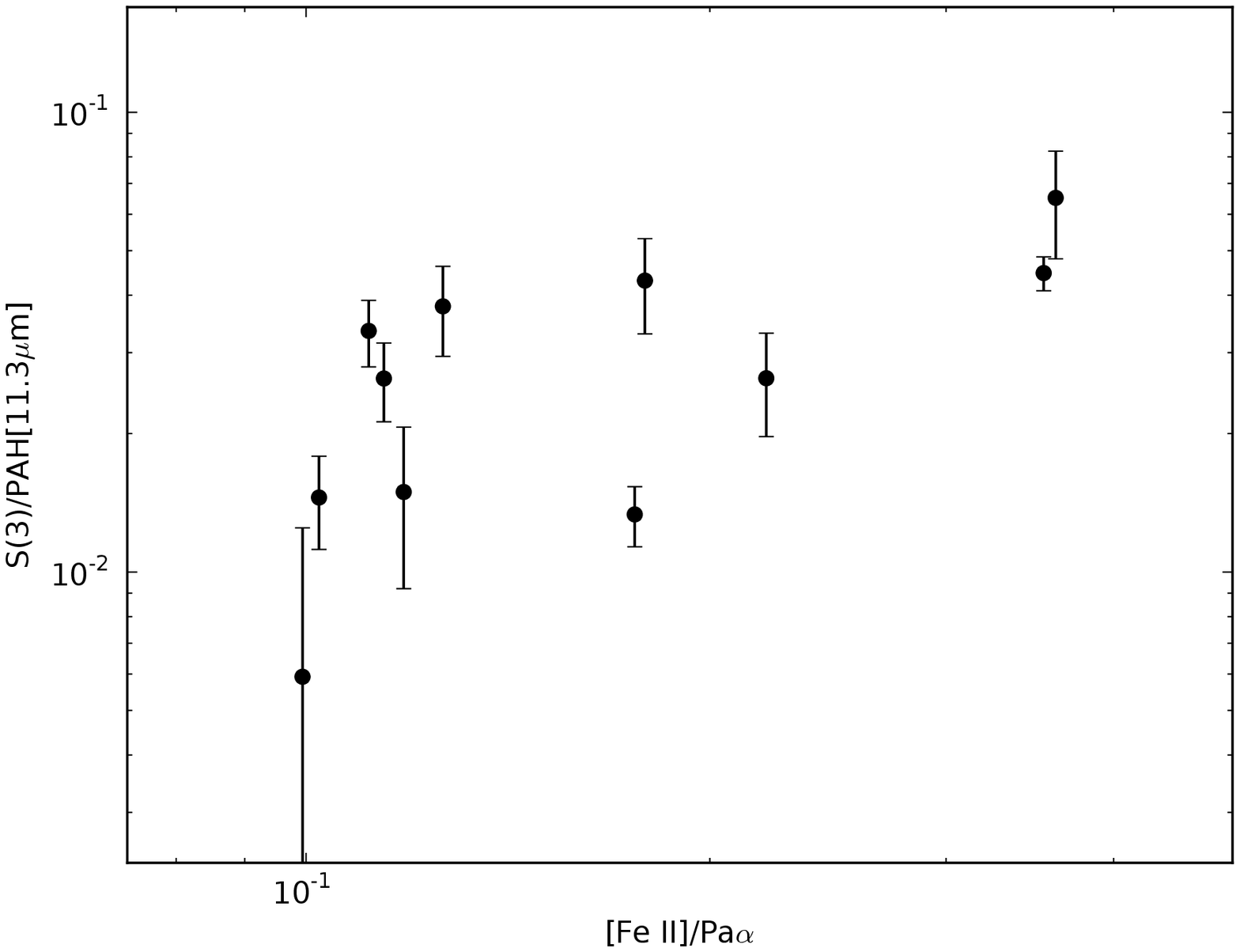}%
\includegraphics[width=80mm]{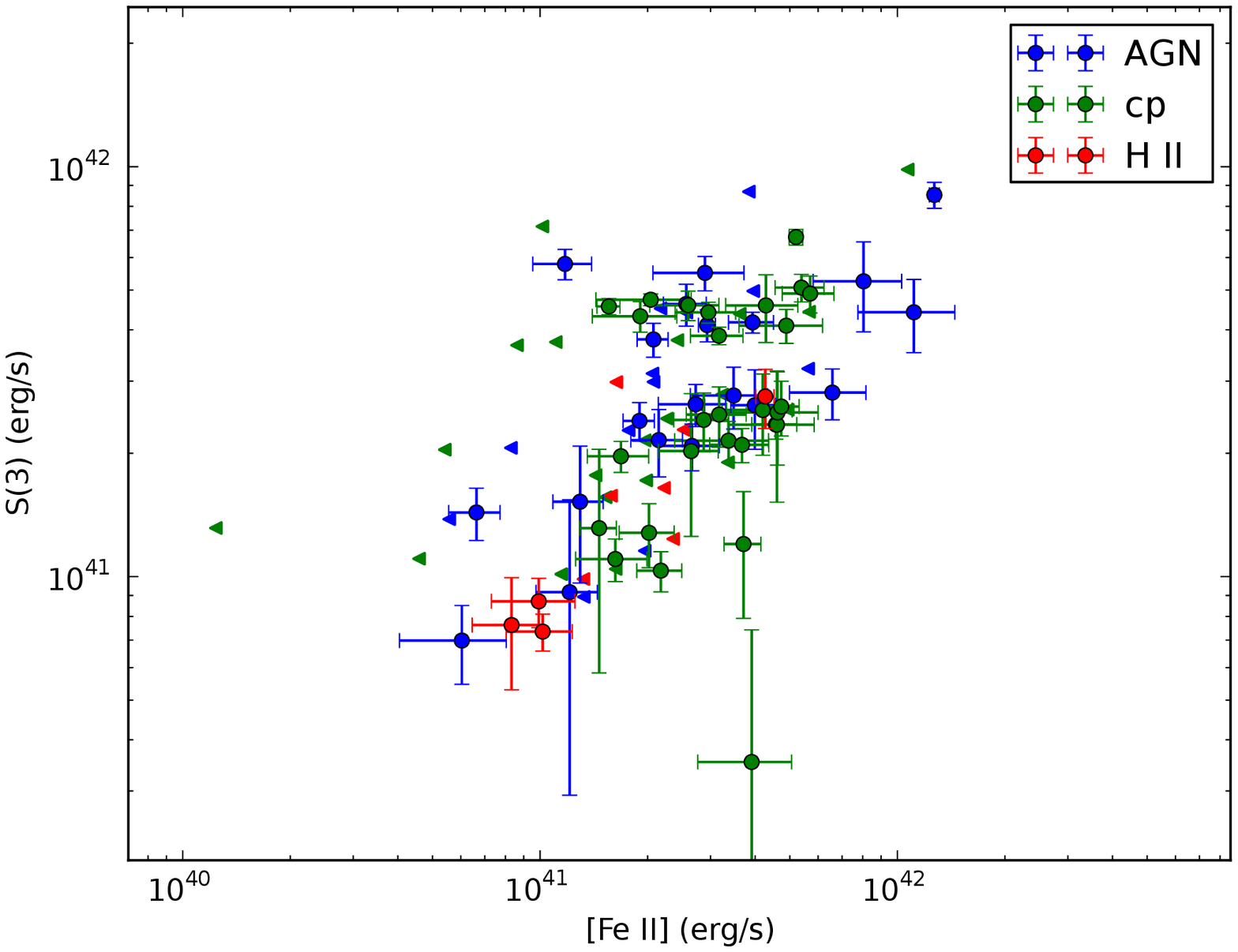}\\
\caption{{\bf Left:} \HPAH and \feiiPa\ are positively correlated ($p_{\rm S}=0.009$), although because of the weakness of \feiiNIR\ the data are available only for 10 objects. {\bf Right:} S(3) and \feiiMIR\ luminosities are linearly proportional to one another. The plot includes 58 \feii\ detections at $>3\sigma$ level (circles) and 44 $3\sigma$ upper limits (triangles). $p_{\rm S}=3.8\times10^{-5}$ for detections, $p_{\rm S}=4.5\times10^{-9}$ for the full set. The fluxes of these features are also strongly linearly correlated; thus, this correlation is robust against the common-distance bias \citep{feig83}. Furthermore, both the flux-flux and the luminosity-luminosity correlations for S(3) and \feiiMIR\ are also recovered using \citet{will11} spectra of the 30 objects that overlap between their sample and ours, with a similar intercept and scatter, and thus the correlation is robust against spectrophotometric uncertainties. }
\label{fig:feii}
\end{figure*}

We find a particularly strong relationship between warm \molh\ and \feii\ emission (Figure \ref{fig:feii}). We use the near-infrared measurements of \feiiNIR\ by \citet{veil97} when available and our own measurements of \feiiMIR\ in the mid-infrared made from the \spi\ spectra. For the 10 objects with \feiiNIR\ measurements we see a tentative correlation between \feiiPa\ and \HPAH. In the mid-infrared, with many more objects we see a much stronger correlation between \feiiMIR\ and S(3) luminosity. Because both line luminosities are derived from the same \spi\ spectra, we do not have to normalize the lines in this case. 

Furthermore, star-forming objects, those dominated by an AGN and composite galaxies all lie on the same \feii-\molh\ sequence, although as we previously established AGNs tend to lie toward its upper end. Strong \feii\ emission is usually associated with shocks driven into a neutral medium, when they destroy dust grains and release iron into the gaseous phase. The \feii\ emission is then excited largely by collisions in the post-shock recombination region \citep{mour00} or by the interstellar radiation field. These conditions are common in supernova remnants: as the effects of many exploding supernovae compound each other, a bubble of hot gas forms and expands as a wind driving shocks into the neutral gas. Near-infrared transitions of \feii\ provide an accurate measure of supernova rate in starburst galaxies \citep{more02,rose12}, and \feiiMIR\ is expected to be as strong as or stronger than the \feiiNIR\ in supernova remnants \citep{reac06}. Thus, the strong correlation between \molh\ and \feii\ in starburst-dominated galaxies confirms that \molh\ in these objects is produced in supernova-driven shocks.

The origin of \feii\ emission in galaxies containing AGNs is more controversial. \citet{moor88} argue that in composite galaxies \feii\ emission is likely to be dominated by supernova-driven shocks in the host galaxy. \citet{mour00} demonstrate that in low-luminosity AGNs, the X-rays from the nuclei dominate excitation of \feii\ over any AGN-driven outflows. However, in our case both \molh\ and \feii\ clearly share similar physical origins, and in the previous section we made the case for AGN-driven winds as the origin of galaxy-wide shocks. We suggest that winds -- whether driven by the starbursts or by the AGNs -- are responsible for shocks that emit both the rotational \molh\ lines and the \feii\ lines in ULIRGs and that these shocked winds dominate over other sources of excitation for the very luminous AGNs and starbursts studied here.

This view is supported by the good correlations between \molh, \feiiMIR\ and \oir\ and the poor correlation between \molh/PAH and \niir, reminiscent of the relationships found by \citep{mour00}. These authors argue that \feii\ and \oi\ are all produced in partly ionized regions, whereas \nii\ comes from fully ionized gas. Therefore, \feii\ and \oi\ are unlikely to be co-spatial with \nii\ and \oiii, explaining the weakness of the correlations between these values. Thus the driving mechanism of the wind is not as important as the physical conditions of the medium it is driven into; as long as the gas is dense and neutral enough, it will yield \oi, \feii\ and apparently \molh\ emission in the post-shock region.

\subsection{Neutral phase}

Na I D absorption is much stronger in ULIRGs than it is in normal star forming galaxies. The median EEW(NaD) is 5.3\AA\ for those of the objects in our sample that have SDSS spectra; for comparison, normal SDSS galaxies with EW(NaD)$>0.8$\AA\ are considered strong Na D absorbers \citep{chen10}. Although some Na D absorption may arise in the atmospheres of cool stars -- this effect contributes up to 80 per cent of the Na D equivalent width in normal galaxies -- the much higher strength of the absorption in ULIRGs implies that the stellar contribution is negligible compared to the interstellar one in our sources.

\begin{figure*}
\centering
\includegraphics[width=80mm]{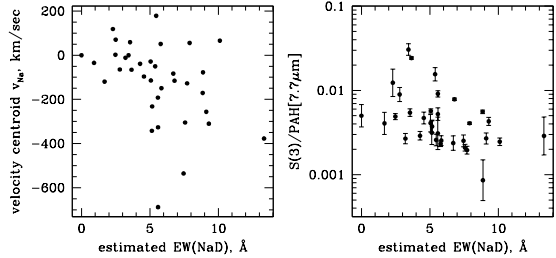}
\caption{{\bf Left:} Objects with strong Na D absorption are more likely to show high Na D blueshifts relative to the galaxy rest-frame ($p_{\rm S}=0.004$). {\bf Right:} Excess \molh\ may be associated with weaker Na D absorption (anti-correlation detected tentatively with $p_{\rm S}=0.08$). }
\label{fig:nad}
\end{figure*}

\citet{chen10} demonstrate that in normal disk galaxies there are two components to the Na D absorption. One is at the redshift of the galaxy; it is associated with the rotationally supported gaseous disk and is more likely to be seen when the galaxy is edge-on. The other one is blue-shifted and associated with the galactic outflows; it is more likely seen when the galaxy is pole-on. The signal-to-noise and the spectral resolution of our data do not allow us to decompose Na D into these two components, but we find an anti-correlation between EEW(NaD) and its velocity centroid $v_{\rm Na}$ ($p_{\rm S}=0.004$; Figure \ref{fig:nad}), indicating that the strongest absorption is associated with the largest blue-shifts. Thus, the strongest Na D absorption in our sample is associated with the outflow component. In contrast to \citet{chen10}, we do not find any correlations between any of the parameters of Na D absorption and galaxy orientation as measured by the apparent ellipticity (taken from \citealt{veil02}). This disagreement can be explained if the starbursts -- and their associated winds -- in ULIRGs are compact and concentrated on scales much smaller than those probed by overall galaxy ellipticities. Alternatively, the correlation may not be detected because we are unable to isolate disk-associated and outflow-associated components separately.

We do not find any strong correlations between \molh, its \texc\ and any of the parameters of the Na D absorption. The strongest trend is between S(3)/PAH[7.7\micron] and EEW(NaD), with $p_{\rm S}=0.08$ (i.e., only tentatively detected; Figure \ref{fig:nad}), in the sense that stronger \molh\ excess is associated with objects with weaker Na D absorption. This is a puzzling and counter-intuitive result that suggests that perhaps the same shocks that lead to excess \molh\ emission eliminate the regions of the interstellar medium where Na D absorption is produced. A larger sample with better quality data for Na D absorption will lead to a more conclusive result on the relationship between the neutral and the warm molecular phases.

In contrast, the kinematics of the ionized gas and the neutral gas phases are unambiguously correlated. Indeed, high outflow velocities in the neutral component are associated with high outflow velocities in the ionized component ($p_{\rm S}\ga 0.001$, depending on the exact kinematic measures used). Thus, in the objects with the largest blue-shifts both Na D absorption and ionized gas emission arise in the same outflow (which presumably consists of pockets of gas with different physical conditions).

Finally, we look at the relationships between various measures of dust extinction. The ones available to us are $\tau[9.7\micron]$ which can be used as a measure of dust obscuration toward the source of the bolometric power; the Balmer ratio H$\alpha$/H$\beta$ which can be used as a measure of dust reddening toward the ionized gas; and EEW(NaD) which measures the column density of cold neutral material toward the stellar component of the galaxy and thus in principle should be an indirect measure of dust absorption toward that source. Surprisingly, although all these values are in principle direct or indirect measures of the column density of dust, no two measures are correlated with one another. This likely reflects the complex geometry of ULIRGs, in which the powering source, the ionized gas and the stellar component might all be distributed over very different physical scales.

We conclude that the three phases of the interstellar medium of ULIRGs that we can probe with our data (warm molecular, cold neutral and ionized) are clearly related to one another, but there is no simple relationship between them. On the one hand, the kinematics of Na D and the [OIII]-emitting ionized gas are correlated. But at the same time, the column density of Na D does not necessarily reflect the amount of dense post-shock gas which is apparently where most of the excess \molh\ emission is produced. Our finding is reminiscent of that by \citet{guil12}. These authors investigated shocks driven into neutral and molecular gas by the impact of a radio jet and found that while shock signatures were clearly present in both media, there was no simple relationship between the kinematics of the two components.

\section{Discussion and conclusions}
\label{sec:conclusions}

In normal star-forming galaxies, the emission from warm (\texc$\sim 100-1000$ K) molecular hydrogen is a minor bi-product of star-formation processes. However, in ULIRGs the tight correlation between \molh\ emission and star-formation indicators breaks down, and these objects show on average several times more \molh\ emission than their star formation rates would suggest. In this paper we use archival \spi\ mid-IR spectra to measure molecular hydrogen rotational emission of 115 ULIRGs in the IRAS 1 Jy sample. We then use a number of morphological and spectral measurements at optical and infrared wavelengths to uncover the source of the excess \molh\ emission. We use PAH luminosities as a star formation indicator and the \molh/PAH ratios to quantify the amount of excess \molh\ emission.

We find a surprising lack of evidence connecting \molh\ emission with merger stages. There is no difference between the \molh/PAH distributions of Pre-Mergers, Mergers, and Old Mergers; nor is there any correlation between \molh/PAH and nuclear separation. This indicates that collision shock is not the primary source of the \molh\ emission we observe. Perhaps the collision velocities of merging galaxies in ULIRG systems are insufficient to create the conditions necessary for the production of excess \molh\ emission. Indeed, the most prominent example of \molh\ excess associated with a collision shock involves a group-like environment and galaxy collision velocities $\sim 1000$ km sec$^{-1}$ \citep{appl06}, presumably much larger than the relative velocities of merging galaxies in our sample (\citealt{coli01} estimate that the typical orbital velocity in ULIRGs in the advanced stage of merger is $\sim 250$ km/sec). 

With collision shocks ruled out as a primary mechanism for \molh\ emission, we investigate other possible sources of \molh\ excess in ULIRGs. We confirm that the primary mechanism of the rotational \molh\ emission is shock excitation by finding a strong correlation between \molh/PAH and optical shock diagnostics, such as \oir. We also find a positive correlation between \molh/PAH and the  widths of optical emission lines. Broadening of optical emission lines beyond the typical line widths associated with galactic rotation has been shown by many authors to be due to galactic outflows; thus, we conclude that excess \molh\ seen in ULIRGs likely arises in outflows as well.

We find a weak correlation between \molh/PAH and the length of tidal tails, especially in Merger-stage galaxies characterized by a single nucleus, but with strong merger signatures. This likely indicates that \molh-producing outflows require some time to develop after the first close passage of the two merging galaxies. The median age of stellar populations in our sample is 250 Myr, consistent with the star-formation time scales in numerical simulations of major mergers. However, unfortunately we do not find any correlations between outflow diagnostics and the age of the stellar population, and thus we cannot pinpoint more precisely the most opportune time for the formation of the \molh-producing outflows.

Furthermore, we find a strong correlation between \molh\ and \feii, an emission feature strongly associated with shocks driven into a largely neutral medium. ULIRGs classified as star-forming galaxies and those containing powerful AGNs all lie on the same \molh-\feii\ sequence. Although the AGN-dominated objects tend to have higher \molh/PAH ratios, the effect of the AGN is fairly subtle (the distributions differ at 90 per cent confidence level, but not at 99 per cent confidence level), and we find it unlikely that direct emission from the AGN is responsible for the excitation of \molh. Instead, we argue that in ULIRGs of all types excess \molh\ emission is produced in strong shocks driven into the neutral interstellar medium either by supernova explosions, or by the radiative pressure of the powerful AGN, or by both. As was previously suggested using observations and numerical simulations \citep{nara08, rupk13a}, we find that the most striking difference between AGN-dominated and star-formation-dominated outflows is the outflow velocity, with maximum blueshifts $v_{02}\sim -1000$ km sec$^{-1}$ for the former and $v_{02}\sim -500$ km sec$^{-1}$ for the latter.

Using our observations, we arrive at the following picture for the origin of the \molh\ excess in ULIRGs. When gas-rich galaxies merge, the gravitational potential of each of them is perturbed by the passing companion. The gas that was initially in the rotational equilibrium within each galaxy disk is now supplied to the centre; at the same time, the potential perturbations lead to the formation of tidal tails. This massive inflow of gas fuels an intense burst of star formation which starts at about the time of the first close approach and can continue at a heightened rate for hundreds of millions of years \citep{rich11, torr12}. The gas is continually supplied, new stars form (and the more massive of them explode as supernovae), and sometimes the conditions are fortuitous for strong AGN activity.

It appears that for the production of \molh\ and \feii\ emission, the exact powering source of outflow (supernova-driven vs AGN-driven) does not matter as much as the physical conditions into which the outflow propagates. Both supernovae and AGNs inject energy into the interstellar medium, leading to a formation of a bubble of hot gas that expands and drives the shocks into the rest of the interstellar medium \citep{ceci02}. Supernovae are active throughout the entire merger, as long as new stars continue to form, whereas the AGN phase tends to last only a short fraction of the merger sequence. However, the instantaneous energy injection by the AGN may be higher leading to higher outflow velocities in this phase \citep{nara08}.

As the outflow propagates into the interstellar medium of the galaxy, it impacts, shocks, entrains and shreds clouds of varying densities and sizes. Thus a wide range of physical conditions is represented within the outflow, including partially ionized regions where \oi\ and \feii\ emission is produced. The strongest correlations that we find are between the \molh\ emission and these emission lines, indicating a strong likelihood that \molh\ is excited in the same regions by supernova-driven and AGN-driven shocks.

\section*{Acknowledgments}

We thank J.E.Greene, L.J.Kewley, N.P.H.Nesvadba, D.A.Neufeld and the anonymous referee for useful discussions, and D.-C.Kim, M.Imanishi and S.Veilleux for making their data electronically available. M.J.H. is supported in part by the Dean's Undergraduate Research Award (JHU) and the Provost's Undergraduate Research Award (JHU). N.L.Z. is supported in part by the Alfred P. Sloan fellowship and Theodore Dunham, Jr. Grant of the Fund for Astrophysical Research. This research has made use of the NASA/IPAC Infrared Science Archive and NASA/IPAC Extragalactic Database, which are operated by the Jet Propulsion Laboratory, California Institute of Technology, under contract with the National Aeronautics and Space Administration. N.L.Z. would like to thank the Institute for Advanced Study (Princeton NJ) for hospitality.  

%\clearpage

\bibliographystyle{mn2e_fix}
\bibliography{master}

\label{lastpage}
\end{document}